\documentclass[11pt]{article}
\def\bSig\mathbf{\Sigma}

\usepackage[margin=1in]{geometry}
\usepackage{authblk}
\usepackage{multicol}
\usepackage{amsmath,tikz}
\usetikzlibrary{calc}
\usetikzlibrary{positioning}

\usepackage[round,comma]{natbib}
\usepackage{amssymb}
\usepackage{url}

\title{Mixture Models for Single-Cell Assays with Application to Vaccine Studies}

\author[1]{Greg Finak}
\author[1]{Andrew McDavid}
\author[3]{Pratip Chattopadhyay}
\author[3]{Maria Dominguez}
\author[1,2]{Steve De Rosa}
\author[3]{Mario Roederer}
\author[1]{Raphael Gottardo}

\affil[1]{Vaccine and Infectious Disease Division, Fred Hutchinson Cancer Research Center (FHCRC), Seattle, WA}
\affil[2]{HIV Vaccine Trials Network, Fred Hutchinson Cancer Research Center (FHCRC), Seattle, WA}
\affil[3]{Vaccine Research Center, NIAID, NIH, 40 Convent Drive, Rm 5509, Bethesda, MD 20892}

\date{\today}                                       
\begin{document}

%\date{{\it Received XXXXX} 2012.}

%\pagerange{\pageref{firstpage}-\pageref{lastpage}} 
%\volume{ }
%\pubyear{ }
%\artmonth{ }

%\doi{12.1234/j.12345-1234.2012.12345.x}

\label{firstpage}
\maketitle
\begin{abstract}
Blood and tissue are composed of many functionally distinct cell subsets. In immunological studies, these can only be measured accurately using single-cell assays. The characterization of these small cell subsets is crucial to decipher system level biological changes. For this reason, an increasing number of studies rely on assays that provide single-cell measurements of multiple genes and proteins from bulk cell samples. A common problem in the analysis of such data is to identify biomarkers (or combinations of biomarkers) that are differentially expressed between two biological conditions (\textit{e.g.}, before/after vaccination), where expression is defined as the proportion of cells expressing that biomarker (or biomarker combination) in the cell subset(s) of interest.
Here, we present a Bayesian hierarchical framework based on a beta-binomial mixture model for testing for differential biomarker expression using single-cell assays. Our model allows the inference to be subject specific, as is typically required when accessing vaccine responses, while borrowing strength across subjects through common prior distributions. We propose two approaches for parameter estimation: an empirical-Bayes approach using an Expectation-Maximization algorithm and a fully Bayesian one based on a Markov chain Monte Carlo algorithm. We compare our method against frequentist approaches for single-cell assays including Fisher's exact test, a likelihood ratio test, and basic log-fold changes. Using several experimental assays measuring proteins or genes at the single-cell level and simulated data, we show that our method has higher sensitivity and specificity than alternative methods. Additional simulations show that our framework is also robust to model misspecification. Finally, we also demonstrate how our approach can be extended to testing multivariate differential expression across multiple biomarker combinations using a Dirichlet-multinomial model and illustrate this multivariate approach using single-cell gene expression data and simulations.
\end{abstract}

%\begin{keywords}
%Mixture modelling, hierarchical modelling, bayesian modelling, single-cell assays, immunology
%\end{keywords}
%\maketitle

\clearpage
\section{Introduction}
\label{s:intro}
Cell populations, particularly in the immune system, are never truly homogeneous; individual cells may be in different biochemical states that define functional but measurable differences between them. 
This single-cell heterogeneity is informative, but lost in assays that measure cell mixtures. 
For this reason, endpoints in vaccine and immunological studies are measured through a variety of assays that provide single-cell measurements of multiple genes and proteins. 
In the 1970s, single-cell analysis was revolutionized with the development of fluorescence-based flow cytometry (FCM). 
Since then, instrumentation and reagent advances have enabled the study of numerous cellular processes via the simultaneous single-cell measurement of multiple surface and intracellular biomarkers (up to 17 biomarkers). 
More recent technological development have drastically extended the capabilities of single-cell cytometry to measure dozens of simultaneous parameters (i.e. proteins, genes, cytokines, etc.) per cell~\citep{Bendall:2011wf}. 
Although cells sorted using well-established surface biomarkers may appear homogeneous, mRNA expression of other genes within these cells can be heterogeneous~\citep{Narsinh:2011gn, Flatz:2011jb} and could further characterize and subset these cells. 
A new technology based on microfluidic arrays combined with multiplexed polymerase chain reactions (PCR) can now be used to perform thousands of PCRs in a single device, enabling simultaneous, high-throughput gene expression measurements at the single-cell level across hundreds of cells and genes~\citep{Pieprzyk:2009uc}. 
While classic gene expression microarrays sum the expression from many individual cells, the intrinsic stochastic nature of biochemical processes results in relatively large cell-to-cell gene expression variability~\citep{vanOudenaarden200915a}. 
This heterogeneity may carry important information, thus single-cell expression data should not be analyzed in the same fashion as cell-population level data. 
Special treatment of single-cell level data, which preserves information about population heterogeneity, is warranted in general. 
For this reason, single-cell assays are an important tool in immunology, providing a functional and phenotypic snapshot of the immune system at a given time. 
These assays typically measure multiple biomarkers simultaneously on individual cells in a heterogeneous mixture such as whole blood or peripheral blood mononuclear cells  (PBMC), and are used for immune monitoring of disease, vaccine research, and diagnosis of haematological malignancies~\citep{Altman:1996wf,Betts:2006dw,Inokuma:2007tn}.

During analysis, cell level biomarker fluorescence intensities are typically thresholded as positive or negative so that subsets with different multivariate $+/-$ combinations can be obtained as Boolean combinations. 
For some assays (\textit{e.g.}, flow cytometry), the positivity thresholds are set based on prior biological knowledge while for others, thresholds are given by the assay technology. 
This is the case for the Fluidigm technology where genes are recorded as absent (not expressed) or present (expressed) at the single-cell level.
 After this thresholding step, we obtain a Boolean matrix of dimension $N\times K$, where $N$ is the number of cells recorded and $K$ is number of biomarkers. 
Using this matrix, one can form $2^K$ putative cell subsets obtained as Boolean combinations. 
When $K$ is large there is a combinatorial explosion of the number of subsets, and many of these might be small or even empty. 
A common statistical problem is, for a given biomarker combination, to identify subjects for whom the proportion of cells expressing that combination is significantly different between two experimental conditions (\textit{e.g.}, before and after vaccination). Note that we use the term `subject' throughout the paper, but the approaches described are general and can be applied to other experimental units (\textit{e.g.}, animal studies).

A motivating example from vaccine research is the flow cytometric intracellular cytokine staining (ICS) assay, which is used to identify and quantify subjects' immune responses to a vaccine. Upon vaccination, antigen in the vaccine is taken up and presented to CD4 or CD8 T-cells via antigen presenting cells.
 While not all T-cells can recognize all antigens, those that recognize antigens in the vaccine become \emph{activated} and produce a variety of cytokines, further promoting the immune response. 
 After activation, this antigen-specific subpopulation proliferates and can persist in the immune system for some time providing \emph{memory} that can more rapidly recognize the same antigen again in the future~\citep{McKinstry:2010ei}. The antigen--specific T-cell subpopulations (i.e. the subset that can respond to one specific antigen) constitute a very small fraction of the total number of CD4 and CD8 T-cells. The ICS assay measures the number of antigen-specific T-cells in PBMC or whole blood by measuring cytokine production in response to activation following stimulation by an antigen that closely matches what was present in the original vaccine. 
Individual cells are labelled using fluorescently conjugated antibodies against phenotypic biomarkers (CD3, CD4, and CD8), used to subset T-cells, and functional biomarkers (cytokines) used to define antigen specific T-cells~\citep{Horton:2007tsa,DeRosa:2004wp,Betts:2006dw}.
A sufficiently large number of cells must be collected to ensure that the rare cell populations can be detected. 
Subsequently, each individual cell is classified as either positive or negative for each maker based on predetermined thresholds, then the number of cells matching each subpopulation phenotype is counted. 

These counts are compared between antigen stimulated and unstimulated samples from a subject to identify significant differences. Subjects who generate a response after stimulation are called \textit{responders}, whereas subjects that do not show any differences are called \textit{non-responders}. 
In many immunological studies, the size of the functionally distinct subpopulations (\textit{i.e.}, the number of positive cells) is very low (relative to the total number of cells), and real biological differences might be difficult to detect. 

Although there is no standard approach to analyzing ICS assays, current methods range from ad-hoc rules based on log-fold changes~\citep{Trigona:2003}, to non--parametric methods~\citep{Sinclair:2004hs}, to permutation tests based on Hotelling's T$^2$ statistics~\citep{Nason:2006dx}, to exact tests of 2x2 contingency tables (\textit{e.g.}, Fisher's exact test and $\chi^2$ test)~\citep{Horton:2007tsa,Proschan:2009ks,Peiperl:2010ej,Nason:2006dx}.
All of these methods test subjects separately, and no information is shared across observations even though one could expect some similarities across responders (or non-responders). %In addition, these methods generally test one marker at a time or pairwise combinations of markers, raising questions of appropriate multiple testing adjustments, or they perform global tests of significance on multiple markers resulting in decreased power to detect small changes in subsets of cytokines~\cite{Proschan:2009ks,Nason:2006dx}\footnote{What are we trying to say here? Let's update this sentence later}. 

The framework developed in this paper, named MIMOSA (Mixture Models for Single Cell Assays), addresses these issues explicitly. In our model, cell counts are modelled by a binomial (or multinomial in the multivariate case) distribution and information is shared across subjects by means of a prior distribution placed on the unknown proportion(s) of the binomial (or multinomial) likelihood. 
In order to discriminate between responders and non-responders, the prior is written as a mixture of two beta (or Dirichlet in the multivariate case) distributions where the hyper-parameters for each mixture component are shared across subjects. 
This sharing of information helps regularize proportion estimates when the cell counts are small, which is typical with single-cell assays, and increases sensitivity and specificity when detecting responders. 
Because our framework is multivariate in nature, multiple cell subsets can be modelled simultaneously, which could help detect small biological changes that are spread out across multiple cell subsets~\citep{Nason:2006dx}. Our paper is organized as follows; Section \ref{s:data} introduces the data and notations used in the paper. In Section \ref{s:DEone}, we present our model for testing differential biomarker expression in the univariate case. Section \ref{s:results} compares our approach to alternative methods and tests the robustness of our model. In Section \ref{s:demarkercombos} we present a multivariate extension of our model that can be used to test multivariate biomarker differential expression and present some results using a single-cell gene expression data. Finally, in Section \ref{s:discussion} we discuss our findings and future work.

\section{Notation and Data} 
\label{s:data}

In the remainder of the paper, we use the following notation to describe our model.  We assume that we observe cell counts from $I$ subjects in two conditions: stimulated and un-stimulated. Each cell can either be positive or negative for a biomarker. Given a set of $K$ biomarkers, the measured cells can be classified into $2^K$ positive/negative biomarker combinations. We denote by $n_{sik}$ and $n_{uik}$, $k=1,\dots, 2^K, i=1,\dots,I$, the observed counts for the $2^K$ combinations in the stimulated and un-stimulated samples, respectively. We denote by $N_{si}=\sum_k n_{sik}$ and $N_{ui}=\sum_k n_{uik}$ the total number of cells measured for subject $i$ in each sample, respectively. For ease of notation, we denote by $\mathbf{y}_i$ the vector of observed counts for subject $i$, \textit{i.e.}, $\mathbf{y}_{i}=(\mathbf{n}_{si}, \mathbf{n}_{ui})$ where $\mathbf{n}_{si}=\{n_{sik}: k=1,\dots,2^K\}$ and $\mathbf{n}_{ui}=\{n_{uik}: k=1,\dots,2^K\}$. Finally, we define $\mathbf{y}=(\mathbf{y}_1,\dots,\mathbf{y}_I)$.

We consider two types of immunological single-cell assays: flow cytometry and single-cell gene expression, as described below.

\textit{Flow cytometry}:
The primary dataset used here is an ICS data set generated as part of a  trial testing the GeoVax DNA and MVA (Modified Vaccinia Ankara) HIV vaccine in a prime-boost regimen (prime at zero and two months, boost at four and six months)~\citep{Goepfert:2011ci}. 
The goal of this data set was to assess the immune response to the vaccine across multiple antigen stimulations, time points, cytokines and T-cell subsets. 
Here, we analyze a subset of the data consisting of 98 subject from the vaccine group at two time points: day 0 and day 182. 
Three cytokines (IFN-$\gamma$, TNF$\alpha$ and IL2) were measured at the single-cell level for each subject and time point, with and without stimulations with an antigen (here we focus on HIV Envelope peptide pool) matching part of the vaccine. For ease of presentation we restricted ourselves to the CD4+ T-cell subsets. Samples on day 0 were taken just before vaccination and no response is expected there. The corresponding samples can be used as negative controls. Conversely, day 182 (26 weeks) should be close to the immunogenicity peak, and many subjects are expected to respond, for some cytokines at least. 

\textit{Fluidigm single-cell gene expression}: This is a single-cell gene expression data set of sorted CD8$+$ T-cells from sixteen subjects. T-cells isolated by flow cytometry from sixteen subjects were stimulated in blocks of four subjects with four different antigens (HIV Gag, HIV Nef, CMV pp65 tm10, and CMV pp65 nlv5) and gene expression post-stimulation measured at the single-cell level using the BioMark system (Fluidigm) $96 \times 96$ well arrays. The expression from the simulated samples  was compared to paired, unstimulated controls.

Although the immunological experiments described above will often look at multiple antigens or stimulations, in the models presented here we consider only one stimulation (i.e. antigen or condition) at a time vs. unstimulated. The issue of multiple antigens is handled through multiple testing correction. 

\section{Differential expression with one biomarker}
\label{s:DEone}
Datasets like the ones presented here are usually analyzed in a univariate fashion to avoid being underpowered due to the large number of combinations and the potential for very small cell counts in many of the combinations. Here, by univariate, we mean that we have only one positive cell subset. This cell subset can be defined by considering the expression of one biomarker alone (marginalizing over all other measured biomarkers) such as A$+$ (\textit{vs.} A$-$), or considering a specific positive biomarker combination (and marginalizing over everything else) such as A$+$ and/or B$+$ (\textit{vs.} A$-$/B$-$). Without loss of generality, we treat the univariate case as a one biomarker case (\textit{i.e.}, $K=1$). In this case, for a given subject, the data can be summarized in a contingency table of $+/-$ cell counts across the un-stimulated and stimulated samples as depicted in Table \ref{tab:twobytwo}.

\begin{table}
\centering
\parbox{0.8\linewidth}{
\caption{2 x 2 contingency table of counts for biomarker positive and negative cells between stimulated ($s$) and unstimulated ($u$) conditions for a given subject $i$.}\label{tab:twobytwo}
\centering
\begin{tabular}{rrr}

  \hline
\multicolumn{1}{l}{}&
\multicolumn{2}{c}{Biomarker}\\
 & Negative & Positive \\ 
  \hline
Stimulated &   $N_{si} - n_{si}$ &   $n_{si}$ \\ 
Unstimulated &   $N_{ui}-n_{ui}$ &   $n_{ui}$ \\ 
   \hline
\end{tabular}
}
\end{table}

For a given subject and stimulation, we consider a biomarker to be differentially expressed if the proportion of positive cells in the stimulated samples is different from the number of positive cells in the un-stimulated sample. Subjects that show differential expression will be called responders for that biomarker. In this section, we are concerned with identifying differential expression one biomarker at a time, using a beta-binomial mixture model as described in what follows. 

\subsection{Beta-binomial model}
\label{s:DE}
For a given subject $i$, the positive cell counts for the stimulated and un-stimulated samples are jointly modeled as follows:

\begin{equation*}
(n_{si}|p_{si}) \sim \mathrm{Bin}(N_{si},p_{si})\quad \text{and}\quad (n_{ui}|p_{ui}) \sim \mathrm{Bin}(N_{ui},p_{ui})\label{eq:bino_likelihood}
\end{equation*}
where $p_{si}$, $p_{ui}$ are the unknown proportions for the stimulated and un-stimulated paired samples, respectively. In order to detect responding subjects, we consider two competing models:
\begin{equation*}
{\cal M}_0: p_{ui}=p_{si}\quad \text{and}\quad {\cal M}_1: p_{ui}\ne p_{si}. \label{eq:models}
\end{equation*}
Under the null model, ${\cal M}_0$, there is no difference between the stimulated and un-stimulated samples, and the proportions are equal (yet the cell counts can differ). Under the alternative model, ${\cal M}_1$, there is a difference in proportions between the two samples and the subject $i$ is a responder. In some studies, such as the ICS data used here, the proportion of positive cells is expected to only increase after stimulation, in which case the alternative model should be defined as $p_s>p_u$. This alternative parametrization is described in Web Appendix B, and we refer to it as the one-sided model.

\subsection{Priors}
\label{ss:priors}
Our model shares information across all subjects using exchangeable~\citep{Bernardo:1996tl} Beta priors on the unknown proportions, as follows: 
 \begin{align*}
(p_{ui} | z_{i}=0)  &\sim \mathrm{Beta}(\alpha_u, \beta_u)\\
(p_{si} | z_{i}=1)  &\sim \mathrm{Beta}(\alpha_s,\beta_s) \quad\mathrm{and}\quad (p_{ui}|z_{i}=1) \sim \mathrm{Beta}(\alpha_u, \beta_u),
 \end{align*}
where $z_i$ is an indicator variable equal to one if subject $i$ is a responder, \textit{i.e.}, ${\cal M}_1$ is true, and zero otherwise, and $\alpha_u, \beta_u, \alpha_s,\beta_s$ are unknown hyper-parameters shared across all subjects. 
Note that the parameters $\alpha_u \text{ and } \beta_u$ are explicitly shared across the two models, whereas $\alpha_s \text{ and } \beta_s$ are only present in the alternative model. 
Finally, we assume that $z_i\sim \mathrm{Be}(w)$ are independent draws from a Bernoulli distribution with probability $w$, where $w$ represents the (unknown) proportion of responders. 
It follows that marginally, \textit{i.e.}, after integrating $z_i$, the $p_{ui}$ and $p_{si}$ are then jointly distributed as a mixture of a one dimensional Beta distribution and a product of two Beta distributions (with a possible constraint), with mixing parameter $w$. 
Treating the $z_i$'s as missing data, the unknown parameter vector $\boldsymbol\theta\equiv(\alpha_u, \beta_u, \alpha_s,\beta_s, w)$ can be estimated in an Empirical-Bayes fashion using Expectation-Maximization algorithm~\citep{Dempster:1977ul} as described in Section \ref{s:estimation}. 
As an alternative, we also describe a fully Bayesian model, where the hyperparameters $\alpha_u, \beta_u$, $\alpha_s,\text{ and }\beta_s$ are each given vague exponential priors with mean $10^3$, and $w$ is assumed to be drawn from a uniform distribution between 0 and 1. 
In this case, all parameters will be estimated via a Markov chain Monte Carlo algorithm as described in Section \ref{s:estimation}. 

\subsection{Parameter estimation}
\label{s:estimation}
In our proposed EM and MCMC algorithms, we greatly simplify our calculations by directly utilizing the marginal likelihoods, $\mathrm{L}_0$ and $\mathrm{L}_1$, obtained after marginalizing $p_{si}$ and $p_{ui}$ from the null and alternative likelihoods. Given the conjugacy of the priors, the marginal likelihoods $\mathrm{L}_0$ and $\mathrm{L}_1$ are available in closed-forms (Web Appendix A), and are given by,

%Our estimation algorithms make direct use of the marginal likelihoods, $\mathrm{L}_0$ and $\mathrm{L}_1$, obtained after integrating out the $p_{\{s,u\}i}$'s for the null and alternative likelihoods, which greatly simplify our calculations. 
 \begin{align*}
 \begin{split}
  	\mathrm{L}_0(\alpha_u,\beta_u|\mathbf{y}_i)
	=&\binom{N_{ui}}{n_{ui}}\binom{N_{si}}{n_{si}}\cdot\\ &\frac{\mathrm{B}(n_{si}+n_{ui}+\alpha_u,N_{si}-n_{si}+N_{ui}-n_{ui}+\beta_u)}{\mathrm{B}(\alpha_u,\beta_u)}
	\end{split}
 \end{align*} 
 
and
\begin{align}
	\begin{split}
\mathrm{L}_1(\alpha_u,\beta_u,\alpha_s,\beta_s|\mathbf{y}_i) 
=&\binom{N_{ui}}{n_{ui}} \binom{N_{si}}{n_{si}}\cdot\\ &\frac{\mathrm{B}(n_{ui}+\alpha_u,N_{ui}-n_{ui}+\beta_u)}{\mathrm{B}(\alpha_u,\beta_u)}\cdot \\ &\frac{\mathrm{B}(n_{si}+\alpha_s,N_{si}-n_{si}+\beta_s)}{\mathrm{B}(\alpha_s,\beta_s)}
\label{model2:unconstrained}
\end{split}
\end{align}
Above, $\mathrm{B}$ is the Beta function. Assuming that the missing data, $z_i, i=1,\dots,I$, are known, we define the complete data log-likelihood:
\begin{equation}
\begin{split}
l(\boldsymbol{\theta}|\mathbf{y},\mathbf{z})=\sum_i z_i l_0(\alpha_u, \beta_u|\mathbf{y}_i) +(1-z_i) l_1(\alpha_u, \beta_u, \alpha_s, \beta_s|\mathbf{y}_i)+\\z_i\log(w)+(1-z_i)\log(1-w),\label{eq:cll}
\end{split}
\end{equation}
where $l_0$ and $l_1$ are the log marginal-likelihoods and $\boldsymbol{\theta}\equiv(\alpha_u, \beta_u, \alpha_s,\beta_s, w)$ is the vector of parameters to be estimated. In the one-sided case, the alternative prior specification must satisfy the constraint $p_s>p_u$, and the marginal likelihood derivation involves the calculation of a normalizing constant that is not available in closed-form but can easily be estimated. All calculations for the one-sided case are described in Web Appendix B. 

\noindent\textbf{EM algorithm}\\
Given an estimate of the model parameter vector $\tilde{\boldsymbol{\theta}}=\left\{\tilde{\alpha}_u,\tilde{\beta}_u,\tilde{\alpha}_s,\tilde{\beta}_s,\tilde{w}\right\}$ and the data $\mathbf{y}$, the E step consists of calculating the posterior probabilities of differential expression, defined by
\[
\begin{split}
&\tilde z_{i} \equiv \mathrm{Pr}(z_i=1|\mathbf{y},\tilde{\boldsymbol{\theta}})=\\ &\frac{\tilde{w} \cdot \mathrm{L}_1(\tilde{\alpha}_u,\tilde{\beta}_u,\tilde{\alpha}_s,\tilde{\beta}_s,|\mathbf{y}_i)}{(1-\tilde{w})\cdot\mathrm{L}_0(\tilde{\alpha}_u,\tilde{\beta}_u|\mathbf{y}_i)+\tilde{w}\cdot\mathrm{L}_1(\tilde{\alpha}_u,\tilde{\beta}_u,\tilde{\alpha}_s,\tilde{\beta}_s|\mathbf{y}_i)}.
\end{split}
\] 
The M-step then consist of optimizing the complete-data log-likelihood over $\boldsymbol{\theta}$ after replacing $z_i$ by $\tilde{z}_{i}$ in \eqref{eq:cll}. Straightforward calculations lead to 
$\tilde w = \sum_i{\tilde{z_i}}/I$, but unfortunately no closed form solutions exist for the remaining parameters. We use numerical optimization as implemented in R's \textit{optim} function to estimate the remaining parameters~\citep{Ihaka:1996ud}.  Starting from some initial values, the EM algorithm iterates between the E and M steps until convergence. In our case, we initialize the $z_{i}$'s using Fisher's exact test to assign each observation to either the null or alternative model components. We then use the estimated $z_i$'s to estimate the $p_{ui}$'s and $p_{si}$'s and use these to set the hyper-parameters to their method-of-moments estimates.

\noindent\textbf{MCMC algorithm}\\
We generated realizations from the posterior distribution via MCMC algorithms~\citep{Gelfand:1996wc}. All updates were done via Metropolis-Hastings sampling except for the $z_i$'s and $w$ that were performed via Gibbs samplings.
Details about the algorithms are given in Web Appendix A. We used the method of \cite{Raftery:1992vp} and \cite{Raftery:1996ws} to determine the number of iterations, based on a short pilot run of the sampler. For each dataset presented here, we calculated that no more than about 1,000,000 iterations with 50,000 burn-in iterations was sufficient to estimate standard posterior quantities. To leave some margin, we used 2,000,000 iterations after 50,000 burn-in iterations for each dataset explored here.

\section{Results}
\label{s:results}
In this section, we apply our MIMOSA model to the data described in Section \ref{s:data}, and present the results of a simulation study based on the ICS data. We evaluated and compared the performance of MIMOSA against Fisher's exact test, the likelihood ratio test, and log fold-change by ROC (receiver operator characteristic) curve analysis and by comparing the \textit{observed FDR} (false discovery rate) against the \textit{nominal FDR} (expected false discovery rate) for each data set~\citep{Storey:2002vj}, where a false discovery (for the ICS data) is a day 0 sample (non--responder) that is incorrectly identified as a responder by the model (or a competing method).

%The constrained model was applied to an ICS data set from a real-world vaccine trial in order to identify responders to antigen stimulation. The unconstrained model was applied to Fluidigm single-cell gene expression data to identify genes differentially expressed between stimulated and unstimulated conditions in populations of single-cells. We also performed simulation studies to assess the performance of the constrained and unconstrained models in a univariate and multivariate settings.

\subsection{ICS}
\begin{figure} %  figure placement: here, top, bottom, or page
   \centering
\begin{tikzpicture} [auto, node distance=0cm]
\node at (0,0) (A){
    \begin{tikzpicture}
    \node[anchor=south west,inner sep=0] at (0,0) (foo) {\includegraphics[width=0.95\columnwidth]{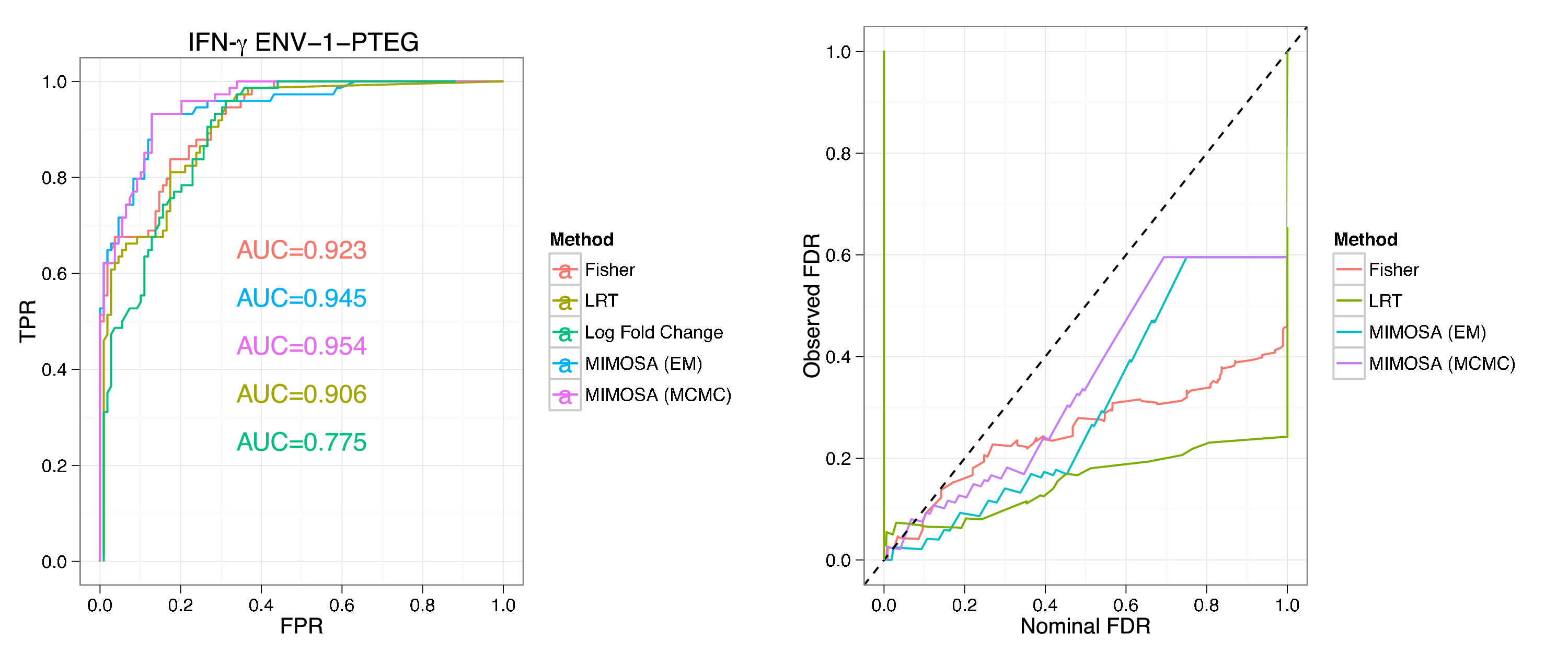}};
    \begin{scope}[x={(foo.south east)},y={(foo.north west)}]
    \node at (0,1) [font=\tiny\sffamily] {A} ;
    \node at (0.5,1) [font=\tiny\sffamily] {B} ;
    \end{scope}
\end{tikzpicture}
 };
 \node [below=of A] (B) {
 \begin{tikzpicture}
    \node[anchor=south west, inner sep=0] at (0,0) (bar){\includegraphics[width=0.95\columnwidth]{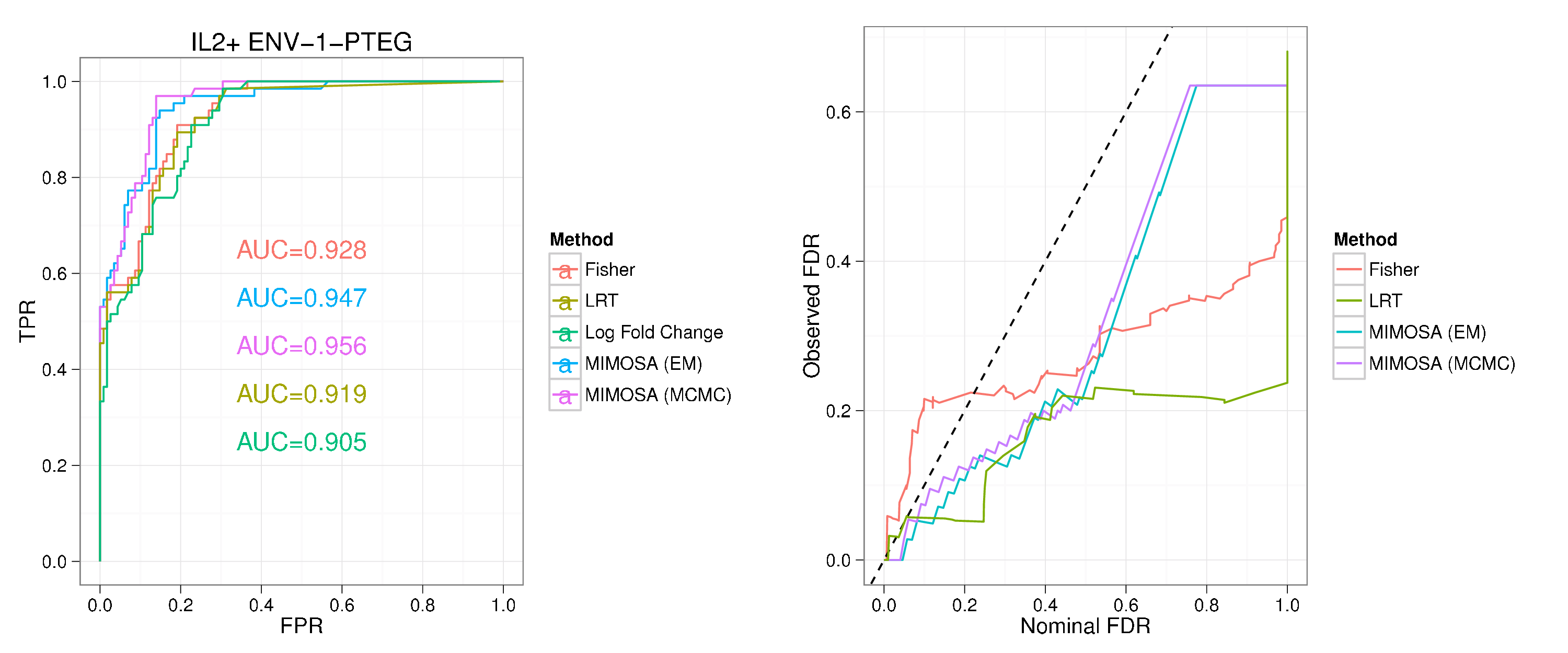}};
    \begin{scope} [x={(bar.south east)},y={(bar.north west)}]
    \node at (0,1) [font=\tiny\sffamily] {C} ;
    \node at (0.5,1) [font=\tiny\sffamily] {D} ;
    \end{scope}
    \end{tikzpicture}
 };
  \node [below=of B] (E) {
 \begin{tikzpicture}
    \node[anchor=south west, inner sep=0] at (0,0) (baz){\includegraphics[width=0.95\columnwidth]{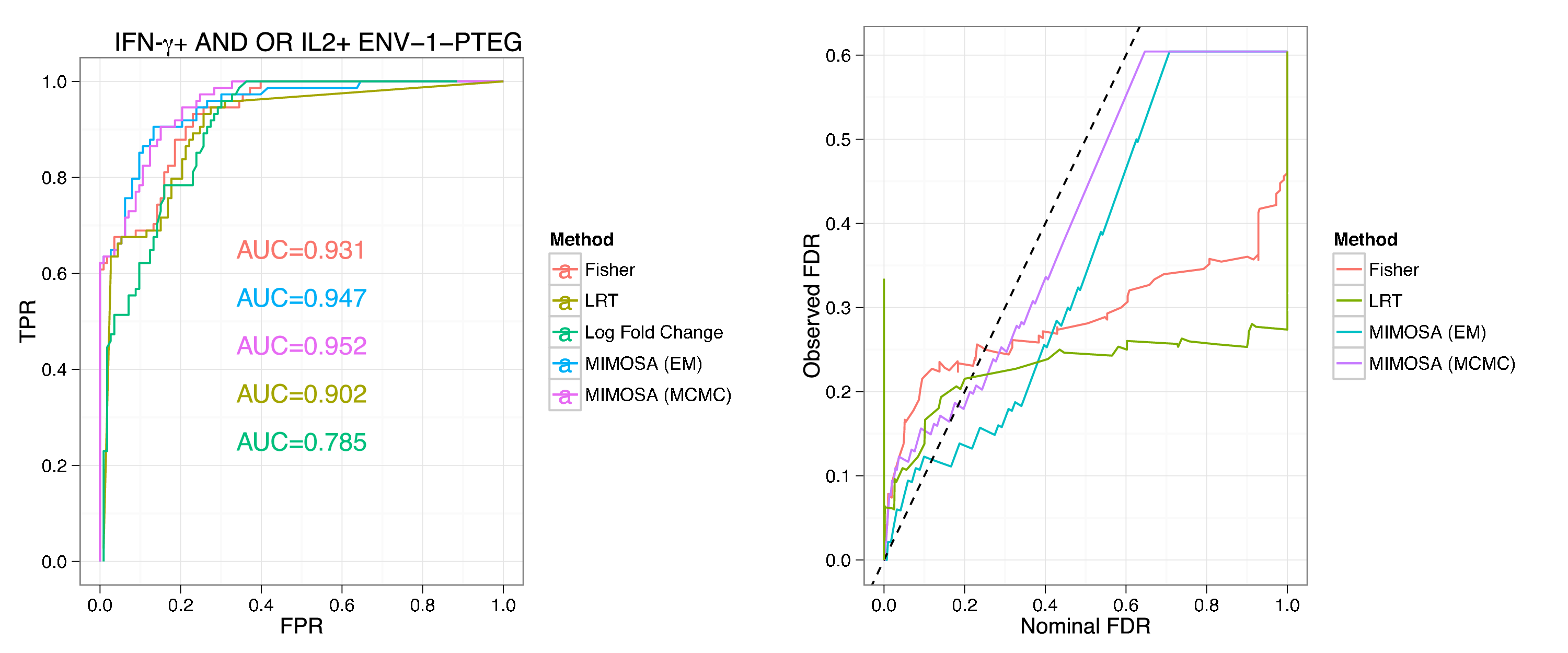}};
    \begin{scope} [x={(baz.south east)},y={(baz.north west)}]
    \node at (0,1) [font=\tiny\sffamily] {E} ;
    \node at (0.5,1) [font=\tiny\sffamily] {F} ;
    \end{scope}
    \end{tikzpicture}
 };
\end{tikzpicture}
   \caption{Performance of MIMOSA (EM and MCMC implementations, one-sided model) and competing methods on ICS data from the example flow cytometry data set. Sensitivity and specificity (ROC analysis) as well as observed and nominal false discovery rates for positivity calls from CD4+ T-cells stimulated with A-B) ENV-1-PTEG and expressing IFN-$\gamma$ or C-D) ENV-1-PTEG and expressing IL2. E-F) ENV-1-PTEG and expressing IFN-$\gamma$ and/or IL2. ROC and FDR plots of other cytokine combinations can be found in Web Figure A.This figure appears in color in the electronic version of this article.}
\label{fig:HVTN065}
\end{figure}
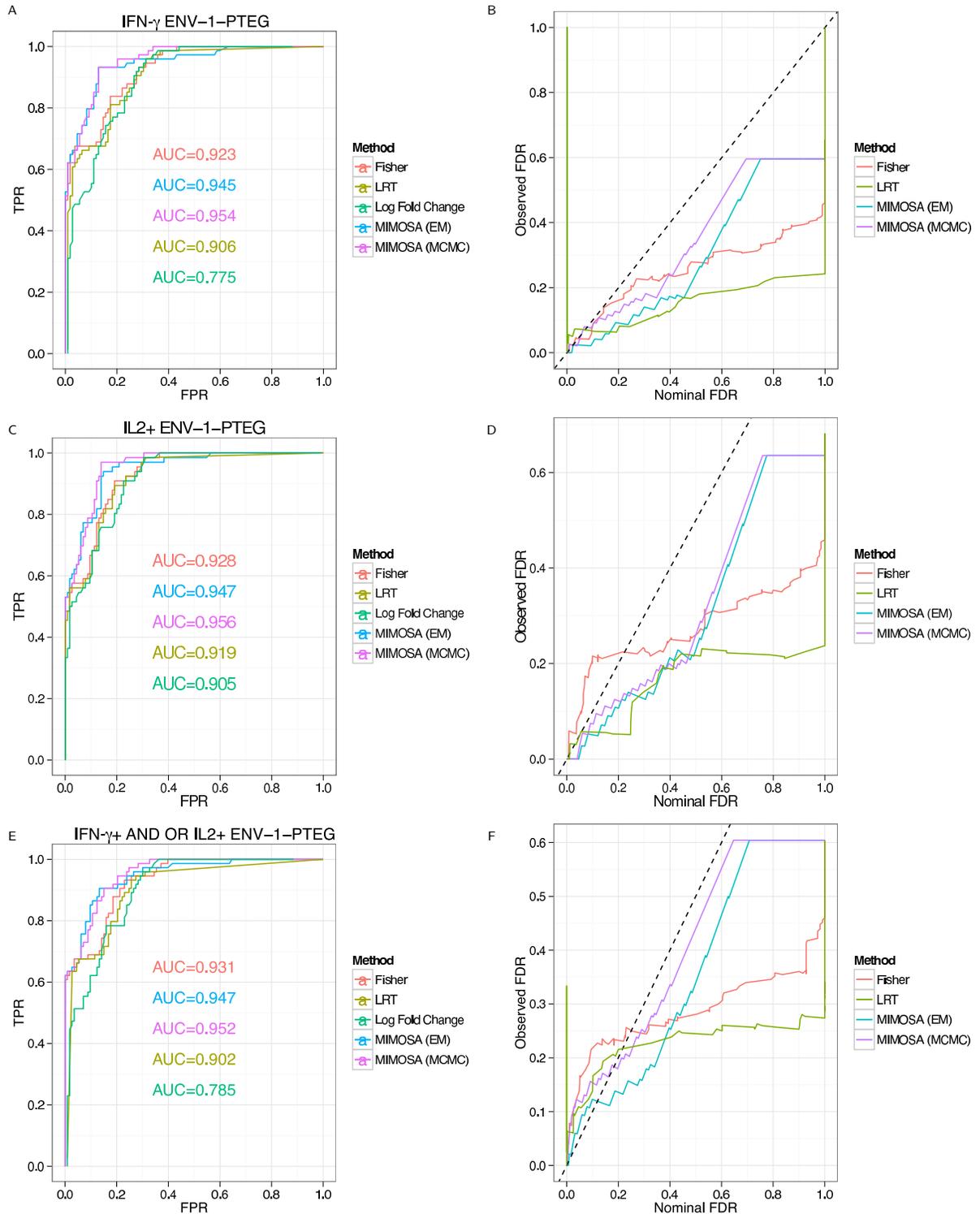

Using the ICS data, we performed an ROC (receiver operator characteristic) analysis to assess the sensitivity and specificity of the one-sided MIMOSA model compared to a one-sided Fisher's exact test, log fold-change, and a likelihood ratio test based on the MIMOSA model for identifying vaccine responders and non-responders. We considered observations at the day 0 time point as true negatives, and observations at the day 182 time point as true positives (potentially underestimating the sensitivity of all methods considered here due to real non-responders at day 182 being treated as true positives). The MIMOSA model has higher sensitivity and specificity than Fisher's exact test, the likelihood ratio test, or log fold-change for discriminating vaccine responders and non-responders  as shown by the ROC curves on Figure~\ref{fig:HVTN065}, panels A,C,E. 
At an FDR between 10-20\%, MIMOSA would lead to about 20\% more true positives being detected. Our comparisons also show that ranking based on log-fold change alone is not reliable and should not be used.
In addition, MIMOSA gave estimates of the observed false discovery rate that are better or comparable to competing methods (Figure~\ref{fig:HVTN065}, panels B,D,F). Here we present the results based on IL2 and IFN-$\gamma$ alone and the subset IL2 and/or IFN-$\gamma$ that were used in the original study~\citep{Goepfert:2011ci}. These results are consistent for other cytokines and cytokine combinations (see Web Figure A).

\subsection{Single-cell gene expression}
\begin{figure}
\centering
\begin{tikzpicture} [auto,node distance=0cm]
\node at (0,0) (A) {
\begin{tikzpicture}
\node[anchor=south west, inner sep=0] at (0,0) (foo) {\includegraphics[width=0.4\columnwidth]{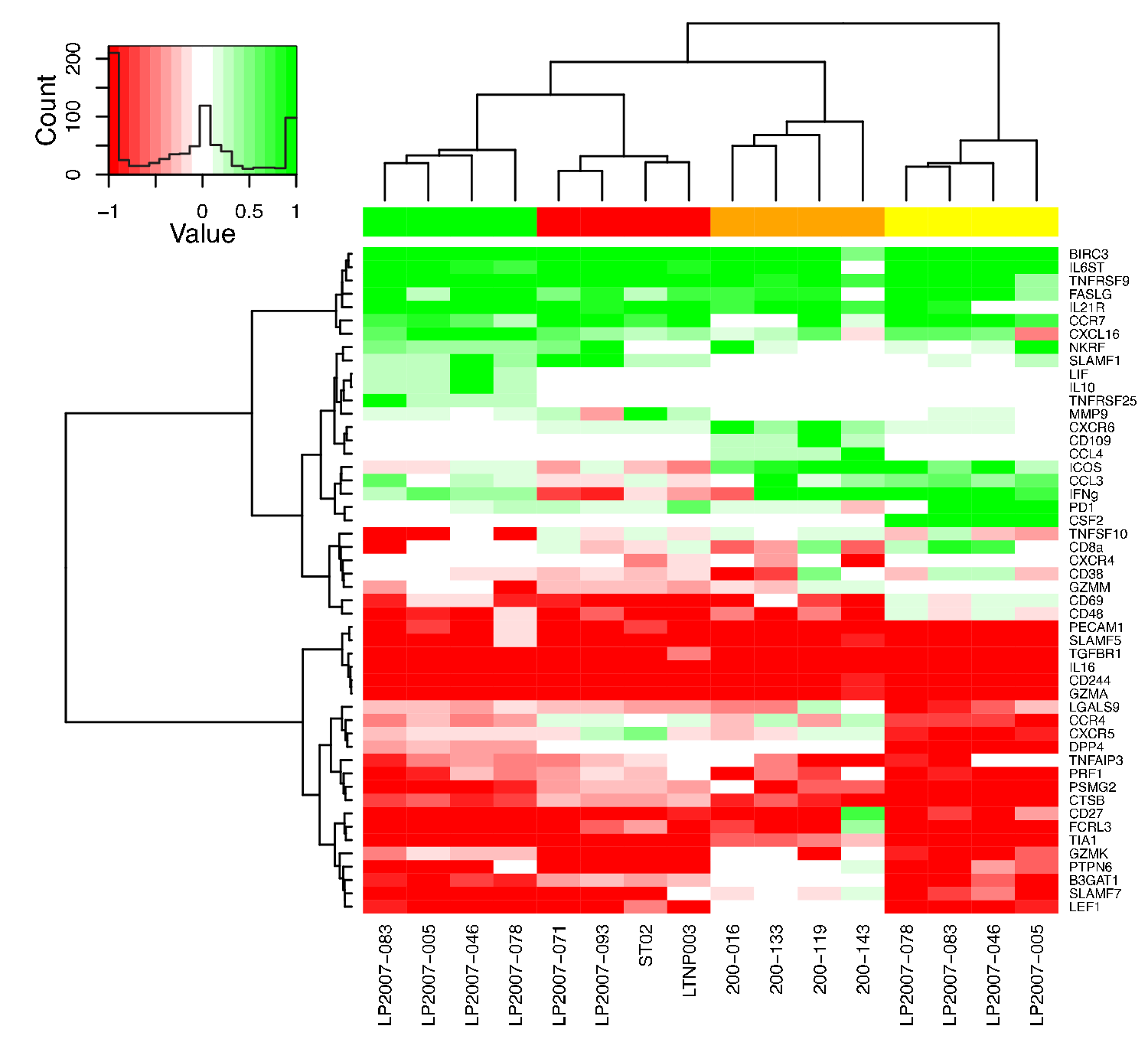}};
\begin{scope} [x={(foo.south east)},y={(foo.north west)}]
\node at (0,1) [font=\tiny\sffamily] {A} ;
\end{scope}
\end{tikzpicture}
};
\node [right=of A] (B) {
\begin{tikzpicture}
\node[anchor=south west, inner sep=0] at (0,0) (bar) {\includegraphics[width=0.4\columnwidth]{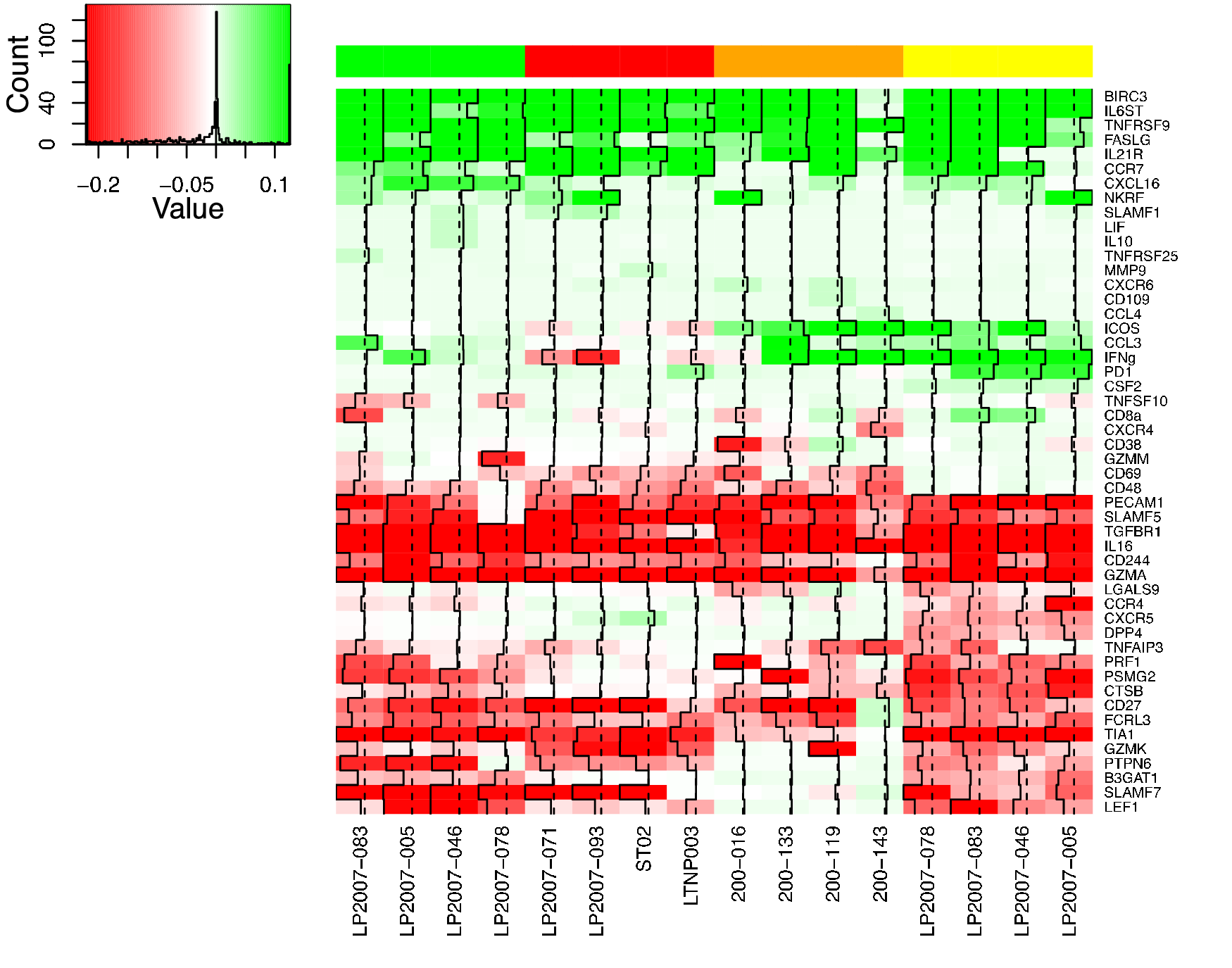}};
\begin{scope} [x={(bar.south east)},y={(bar.north west)}]
\node at (-0.1,1.1) [font=\tiny\sffamily] {B} ;
\end{scope}
\end{tikzpicture}
};
\node [below=of A] (C) {
\begin{tikzpicture}
\node[anchor=south west, inner sep=0] at (0,0) (baz) {\includegraphics[width=0.4\columnwidth]{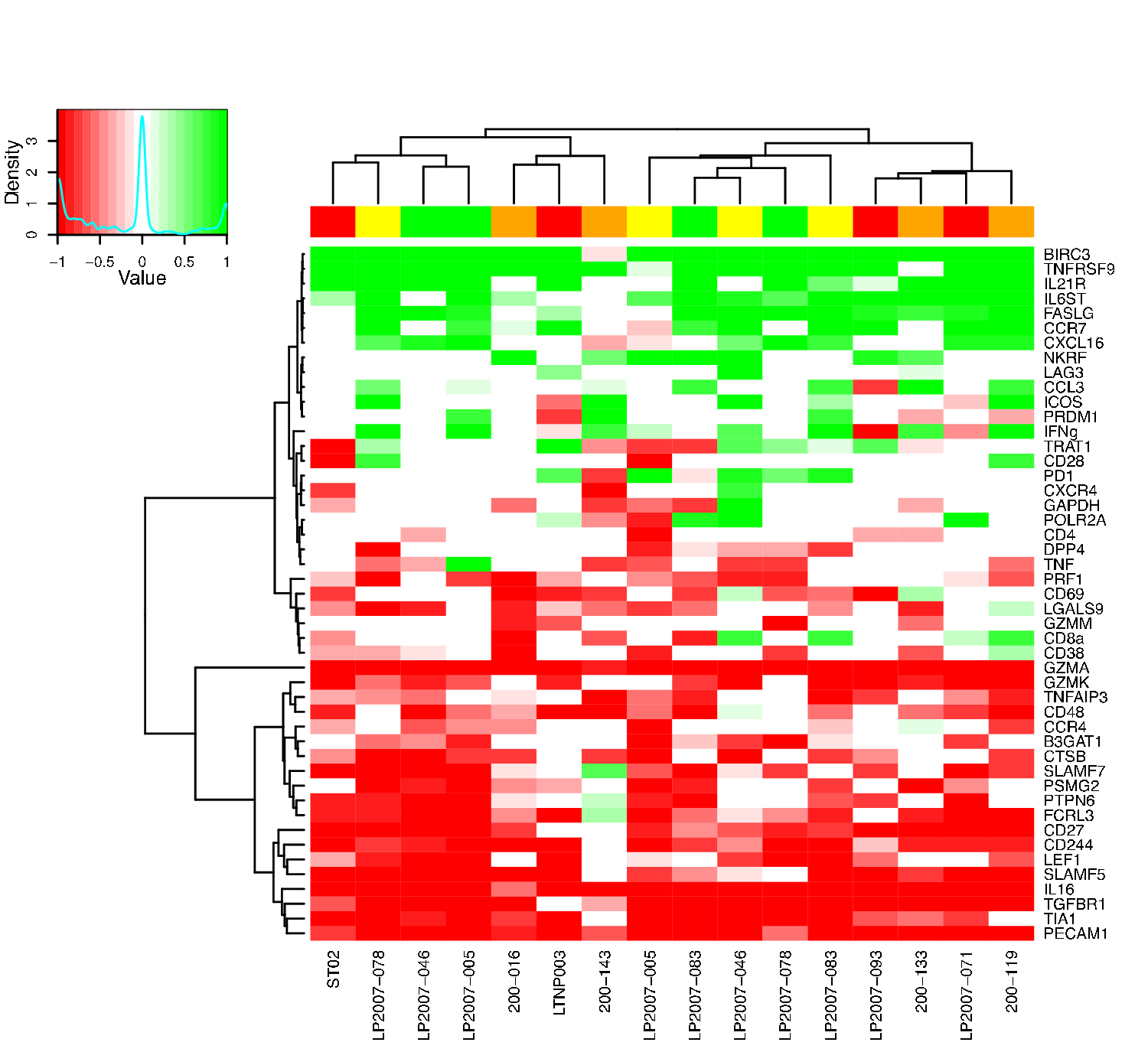}};
\begin{scope} [x={(baz.south east)},y={(baz.north west)}]
\node at (-0.02,1) [font=\tiny\sffamily] {C} ;
\end{scope}
\end{tikzpicture}
};
\end{tikzpicture}
\caption{Signed posterior probability, difference and log-odds ratio of the proportion of single-cells expressing each gene on a 96x96 Fluidigm array. The posterior probability of response times the sign of the change in expression is shown in A) (red indicates a decrease, green an increase, relative to the control). Columns and rows are clustered based on these signed posterior probabilities. B) The posterior differences in proportion of cells expressing a gene in the stimulated vs. control samples. Rows and columns are ordered as in A) for comparison. The traces show the deviations of each cell from zero. Colors along the columns denote different stimulations (green: CMV pp65 nlv5, red: HIV Gag, orange: HIV Nef, yellow: CMV pp65 tm10). C) Clustering of the signed q-values from Fisher's exact test. Genes selected from Fisher's exact test at the 10\% FDR level. This figure appears in color in the electronic version of this article.}
\label{fig:fluidigm}
\end{figure}
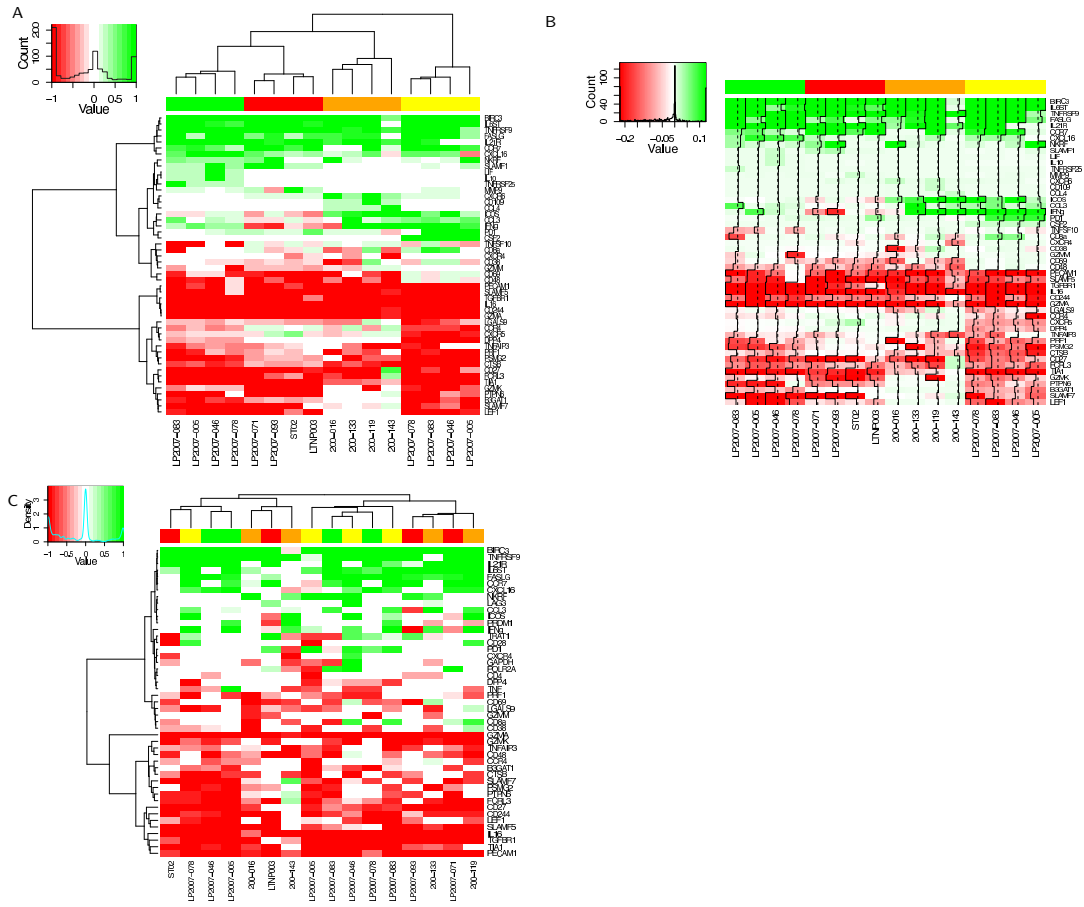

We applied the MIMOSA model to a Fluidigm single-cell gene expression data set. We used the two-sided MIMOSA model because genes could be regulated upward or downward upon stimulation. In order to detect stimulation specific changes of expression, we fit our model to each gene within each stimulation. The results presented in Figure~\ref{fig:fluidigm} show that MIMOSA identifies stimulation-specific differences in the proportions of cells expressing each gene while preserving inter-subject variability (Figure~\ref{fig:fluidigm} A,B). These patterns are evident in the  posterior probabilities (Figure~\ref{fig:fluidigm} A) and preserved in the posterior estimates of the differences of proportions (Figure~\ref{fig:fluidigm} B). A similar analysis using a two-sided Fisher's exact test and clustering the signed FDR adjust p-values (Figure~\ref{fig:fluidigm} C) does not reveal any stimulation-specific patterns. For an FDR of 10\%, Fisher's exact test identified 47 significant genes, while MIMOSA identified 50 significant genes. Both methods identified 39 genes in common.

\subsection{Simulation Studies}

%Simulation studies were based on hyper-parameter estimates from a dataset is from a phase-I (safety and efficacy) trial of an adenoviral vector vaccine in individuals without prior immunity, measuring four cytokines via intracellular cytokine staining (ICS) in two cell populations from 20 individuals at two time points (zero and 28 days post-vaccination, see supplementary material~\ref{supp:statpublished} for details)~\cite{Peiperl:2010ej}. The statistical analysis of ICS data in the published trial is described in the original manuscript and outlined in the supplementary information (supplementary material~\ref{supp:statpublished})~\cite{Peiperl:2010ej}. The goal of this data set was to assess and quantify response rates of CD4 and CD8 T-cell populations to different antigens.

\begin{figure} %  figure placement: here, top, bottom, or page
   \centering
   \begin{tikzpicture} [auto, node distance=0cm]
   \node at (0,0) (A) {
\begin{tikzpicture}
    \node [anchor=south west] at (0,0) (foo) {\includegraphics[width=0.5\columnwidth]{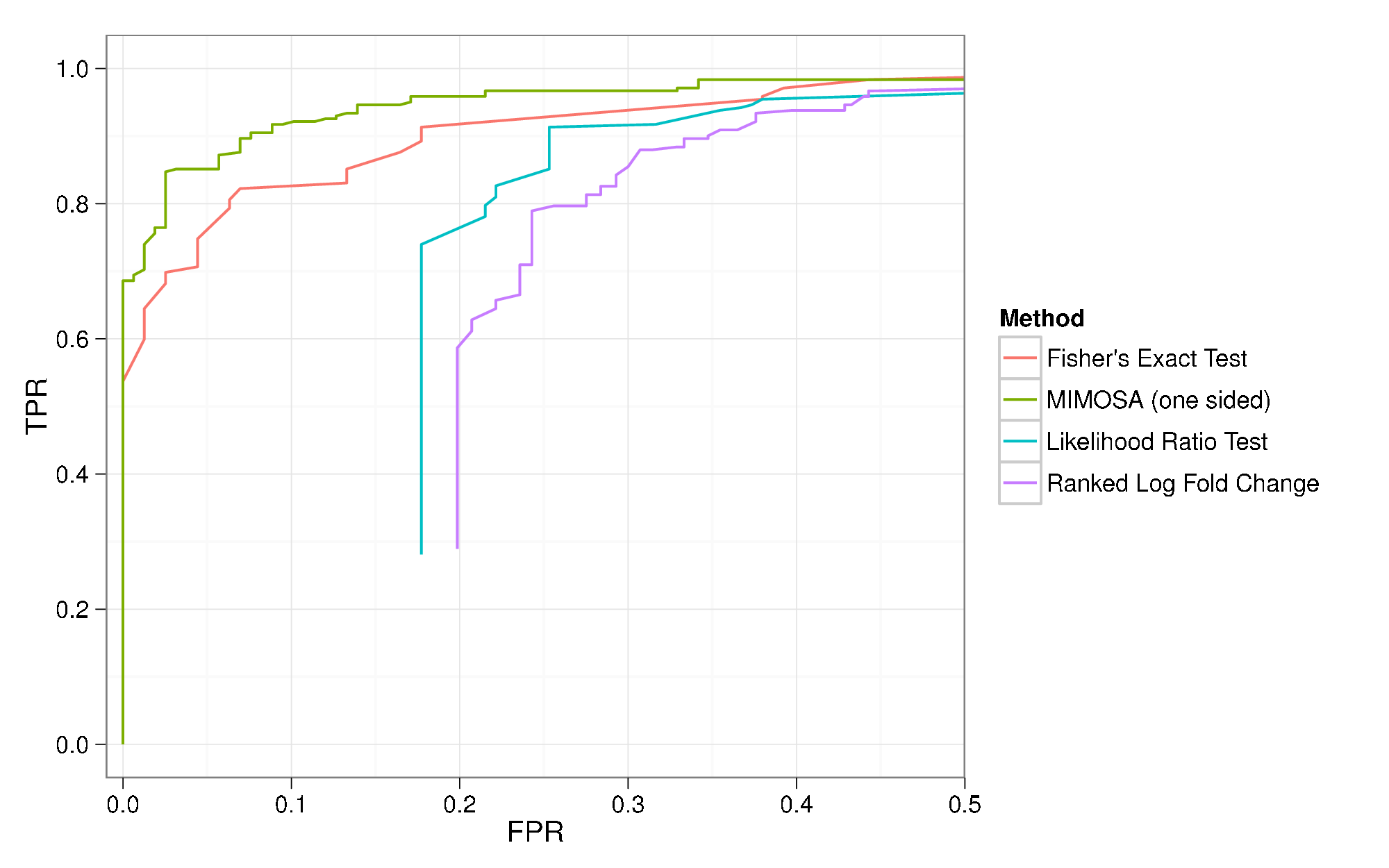}};
    \begin{scope}[x={(foo.south east)},y={(foo.north west)}]
        \node at (0,1) [font=\tiny\sffamily] {A} ;
        \end{scope}
    \end{tikzpicture}
    };
    \node [right=of A] (B) {
    \begin{tikzpicture}
    \node [anchor=south west] at (0,0) (bar) {\includegraphics[width=0.5\columnwidth]{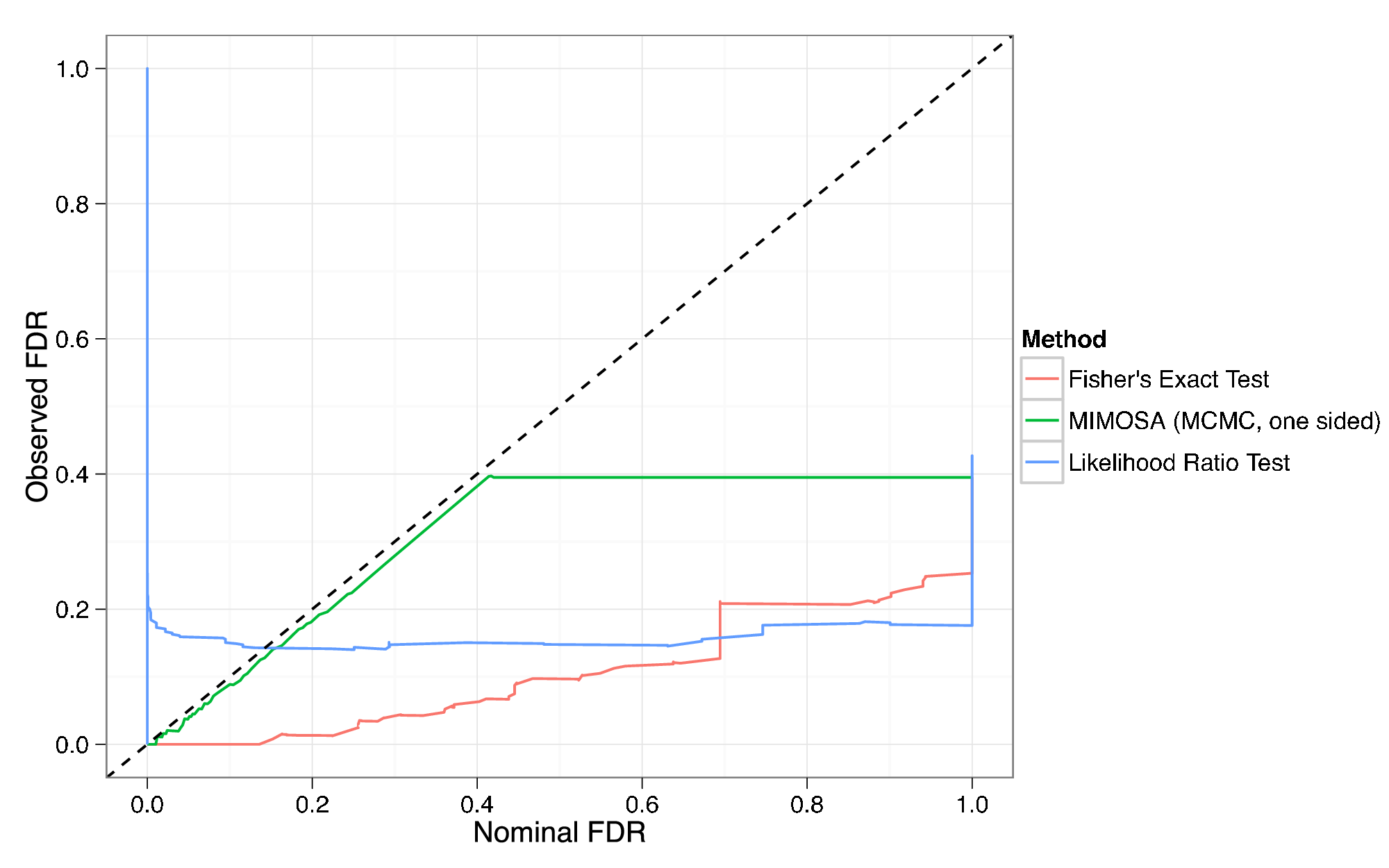}};
    \begin{scope}[x={(bar.south east)},y={(bar.north west)}]
    \node at (0,1) [font=\tiny\sffamily] {B} ;
    \end{scope}
    \end{tikzpicture}
    };
        \node [below=of A] (C) {
    \begin{tikzpicture}
    \node [anchor=south west] at (0,0) (c) {\includegraphics[width=0.5\columnwidth]{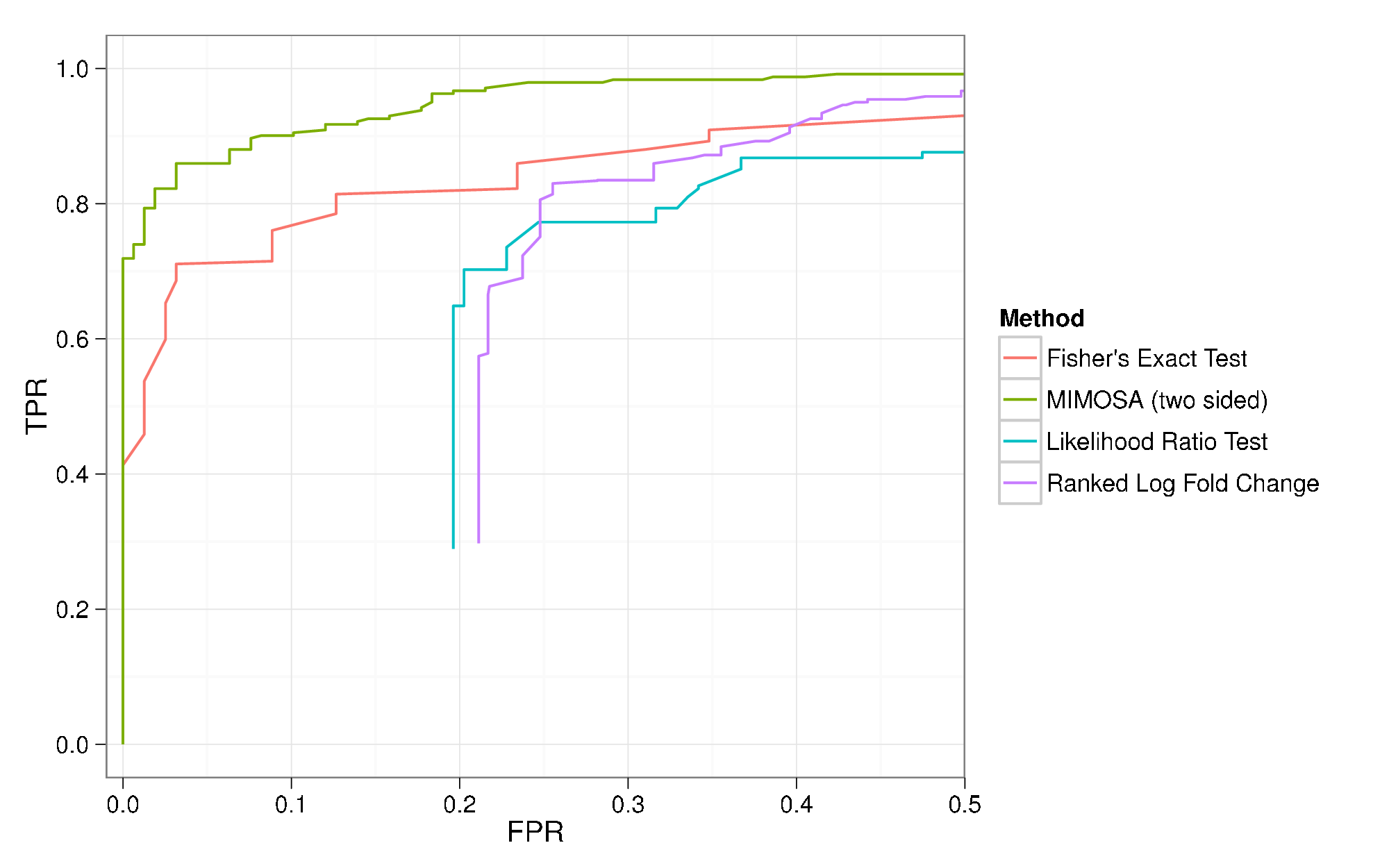}};
    \begin{scope}[x={(c.south east)},y={(c.north west)}]
    \node at (0,1) [font=\tiny\sffamily] {C} ;
    \end{scope}
    \end{tikzpicture}
    };
    \node [right=of C] (D) {
    \begin{tikzpicture}
    \node [anchor=south west] at (0,0) (d) {\includegraphics[width=0.5\columnwidth]{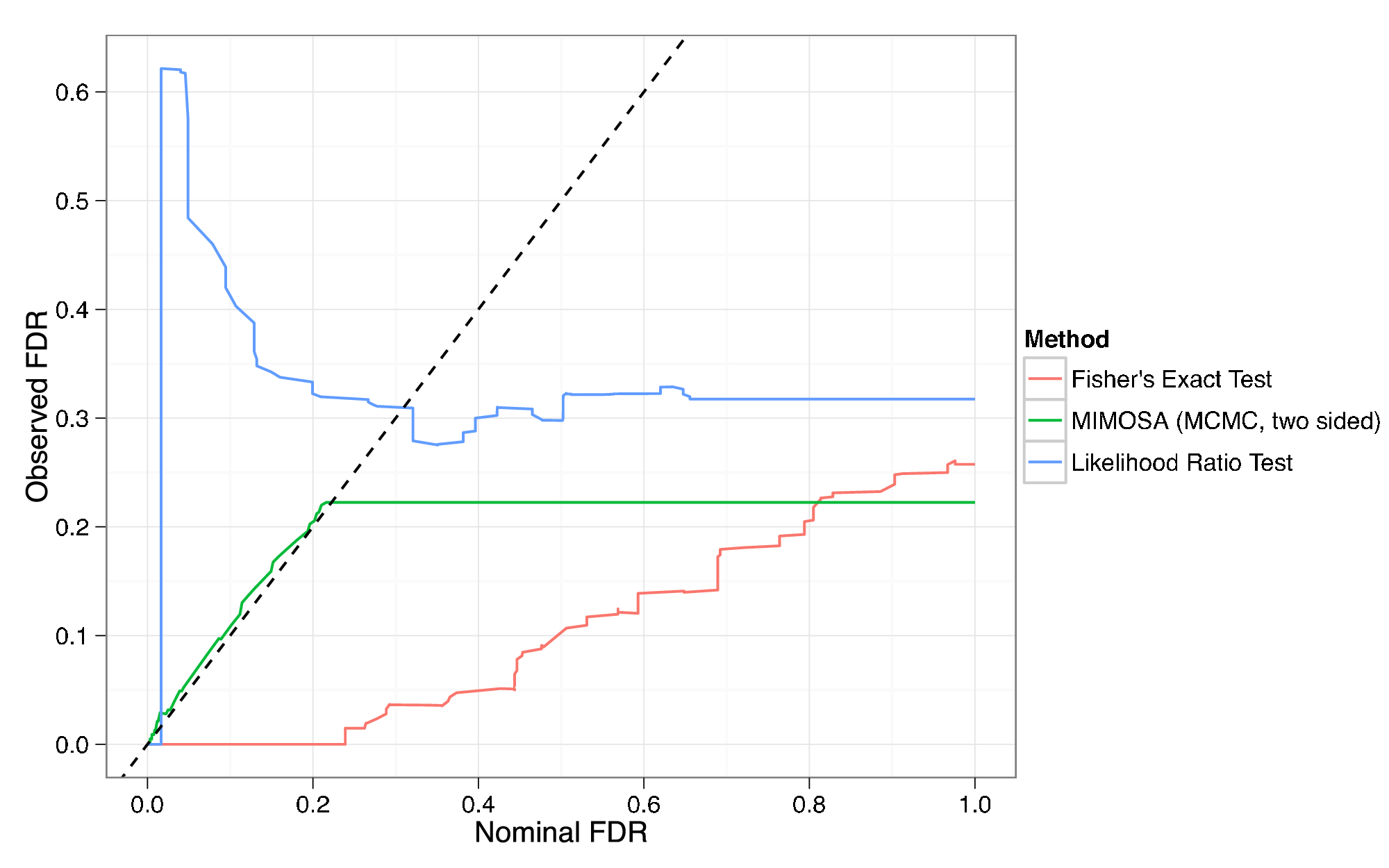}};
    \begin{scope}[x={(d.south east)},y={(d.north west)}]
    \node at (0,1) [font=\tiny\sffamily] {D} ;
    \end{scope}
    \end{tikzpicture}
    };

\end{tikzpicture}
   \caption{Comparison of positivity detection methods on data simulated from the one-sided and two--sided models. Ten simulations were generated at an $N$ of 5,000 total counts using hyper-parameter estimates from real ICS data (IFN-$\gamma$ expressing CD4+ T-cells stimulated with ENV-1-PTEG from HVTN065) with a five-fold effect size between responder and non-responder components. A) Average ROC curve over the 10 simulated data sets (N=5,000), one--sided B) Average observed and nominal false discovery rate over 10 simulated data sets (N=5,000), one--sided. C) Average ROC curves, two--sided model. D) Average observed and nominal FDR, two--sided model. Curves are shown for MIMOSA, Fisher's exact test, the likelihood ratio test, and log fold-change. Results for MIMOSA fit to a model violating model assumptions, as well as other values of N are in Web Figure B. This figure appears in color in the electronic version of this article.}\label{fig:simulations}
\end{figure}

We examined the performance of the constrained ($p_s>p_u$) and unconstrained ($p_s \ne p_u$) beta-binomial mixture models via simulations. For the simulation, we used hyper--parameters estimated from a one-sided MIMOSA model fit to ICS data (IL2 univariate) from the primary immunogenicity time point. We simulated data from this constrained model with 200 observations, a response rate of 60\%, $N=$ 1,000, 5,000, and 10,000 events, with ten independent realizations of data for each $N$. We fit the one-sided MIMOSA model to this data. We evaluated the sensitivity and specificity of the model's ability to correctly identify observations from the ``responder'' and ``non-responder'' groups through analysis of ROC curves, and compared against Fisher's exact test, the likelihood ratio test, and log fold-change. We repeated this procedure for the two-sided models fit to two-sided data (Figure~\ref{fig:simulations} A,C). In addition, we examined the nominal \textit{vs.} observed FDR to assess the ability of each method to properly estimate the FDR (Figure~\ref{fig:simulations} B,D). 

For both the constrained and unconstrained simulations, MIMOSA was superior to competing methods, including Fisher's exact test, with respect to sensitivity and specificity even at small values of $N$ (Figure~\ref{fig:simulations} A and C and Web Figure B, panel E). Additionally, the estimated FDR for MIMOSA more closely reflected the nominal FDR compared to Fisher's exact test and competing methods (Figure~\ref{fig:simulations}, panels B, D, and Web Figure B panel F). 
%Since the constrained model relies on Monte-Carlo integration to estimate the normalizing constant in the likelihood calculation, which can be computationally costly, we approximated the.

To assess the sensitivity of the model to deviations from model assumptions, we repeated the simulations with the cell proportions drawn from  truncated normal distributions with support $(0,1)$, rather than beta distributions. The means and variances of the truncated normal distributions were set to the maximum likelihood estimates of the beta distributions defined by the hyper--parameters  $\alpha$ and $\beta$ estimated from the ICS data set (see Web Figure B panels C and D). Even under these departures from the model assumptions, the unconstrained MIMOSA model outperformed Fisher's exact test.

\section{Differential expression across biomarker combinations}
\label{s:demarkercombos}
Our beta-binomial model described in Section \ref{s:DE} can be generalized to a Dirichlet-multinomial model to assess differential expression across multiple biomarker combinations. As described in the data section, we now have counts for each biomarker combination, denoted by  $\mathbf{n}_{si}=\{n_{sik}: k=1,\dots,2^K\}$ and $\mathbf{n}_{ui}=\{n_{uik}: k=1,\dots,2^K\}$.
%Table~\ref{tab:multdir} gives an illustration of the counts with two putative markers A and B.
\subsection{Model}

In our multivariate model, the beta distribution is replaced by a multinomial distribution, as follows:
\begin{equation*}
 (\mathbf{n}_{ui}|\mathbf{p}_{ui},) \sim \mathcal{M}(N_{ui},\mathbf{p}_{ui})\quad\text{and}\quad (\mathbf{n}_{si}|\mathbf{p}_{si}) \sim \mathcal{M}(N_{si},\mathbf{p}_{si})\label{eq:mult_likeliehood}
 \end{equation*}
%; \text{if $i$ is a non-responder}\\
% &(\mathbf{n}_{ui}|\mathbf{p}_{ui}) \sim \mathcal{M}(\mathbf{p}_{ui},N_{ui}); (\mathbf{n}_{si}|\mathbf{p}_{si}) \sim \mathcal{M}(\mathbf{p}_{si},N_{si});\text{if $i$ is a responder}
%\end{align} 
where $N_{\{s,u\}i}=\sum\limits_{k=1}\limits^{2^K} n_{\{s,u\}ik}$ are the number of cells collected and $\mathbf{p}_{ui}$ and $\mathbf{p}_{si}$ are the unknown proportions for the un-stimulated and stimulated samples, respectively.

\subsection{Prior}
As in the one-biomarker case, we share information across subjects using an exchangeable prior on the unknown proportions. This time the beta priors are replaced by Dirichlet priors, such that
\begin{align*}
(\mathbf{p}_{ui}|z_i=0) &\sim \mathrm{Dir}(\boldsymbol{\alpha}_u),\\\nonumber
(\mathbf{p}_{ui}|z_i=1) &\sim \mathrm{Dir}(\boldsymbol{\alpha}_u) \quad \text{and}\quad (\mathbf{p}_{si}|z_i=1) \sim \mathrm{Dir}(\boldsymbol{\alpha}_s),%\label{eq:dir_prior}
\end{align*}
where the indicator variable $z_i$ is defined in Section \ref{ss:priors}, \textit{i.e.}, $z_i\sim\mathrm{Be}(w)$. As in the beta-binomial case, both an EM and MCMC algorithms can be used for parameter estimation. When using a fully Bayesian approach via MCMC, we use the same priors for $\boldsymbol{\alpha}_{\{u,s\}}$ and $w$ as for the beta-binomial model. 

\subsection{Parameter estimation}
Again, to simplify the estimation problem, we make use of the marginal likelihoods that can be obtained in closed forms (see Web Appendix C). For the null component, the marginal likelihood $\mathrm{L}_0$ is given by,
\begin{align*}
\mathrm{L}_0(\boldsymbol{\alpha}_u|\mathbf{n}_{si},\mathbf{n}_{ui}) &= \frac{ \mathrm{B}(\boldsymbol{\alpha}_{u}+\mathbf{n}_{ui}+\mathbf{n}_{si})}{\mathrm{B}(\boldsymbol{\alpha}_u)} \cdot \frac{N_{si}!}{\prod_k n_{sik}!} \cdot \frac{N_{ui}!}{\prod_k n_{uik}!},
\end{align*}
where $\mathrm{B}$ is the $2^K$-dimensional Beta function defined as $\mathrm{B}(\boldsymbol{\alpha})=\prod_k\Gamma(\alpha_k)/\Gamma(\sum_k\alpha_k)$. Similarly the marginal likelihood for the alternative model is given by 
\[
\mathrm{L}_1(\boldsymbol{\alpha}_u,\boldsymbol{\alpha}_{si}|\mathbf{n}_{si},\mathbf{n}_{ui}) =\frac{\mathrm{B}(\boldsymbol{\alpha}_{u}+\mathbf{n}_{ui}) \mathrm{B}(\boldsymbol{\alpha}_{s}+\mathbf{n}_{si})}{\mathrm{B}(\boldsymbol{\alpha}_s)\mathrm{B}(\boldsymbol{\alpha}_u)} \cdot \frac{N_{si}!}{\prod_k n_{sik}!} \frac{N_{ui}!}{\prod_k n_{uik}!}.%\label{eq:postmult}
\]
The estimation procedures (both EM and MCMC based) for the Dirichlet--multinomial distribution are the same as for the beta-binomial model except that the number of parameters to estimate is larger. We initialize the $z_i$ in the EM algorithm with the positivity calls from the multivariate Fisher's exact test. In our experience, the performance of the EM algorithm greatly deteriorates for $K>3$, and is more dependent on the initial values and can fail to converge in many instances. Although our MCMC algorithm is slightly more computational, it does not suffer from this problem and provides a robust alternative when $K$ is large. More details about our multivariate MCMC algorithm is given in Web Appendix C.

\subsection{Polyfunctionality in Fluidigm Single-Cell Gene Expression Data}
\begin{figure}
\centering
\begin{tikzpicture} [auto, node distance=0cm]
\node at (0,0) (A) {
\begin{tikzpicture}
\node[anchor=south west, inner sep=0] at (0,0) (foo) {\includegraphics[width=0.4\columnwidth]{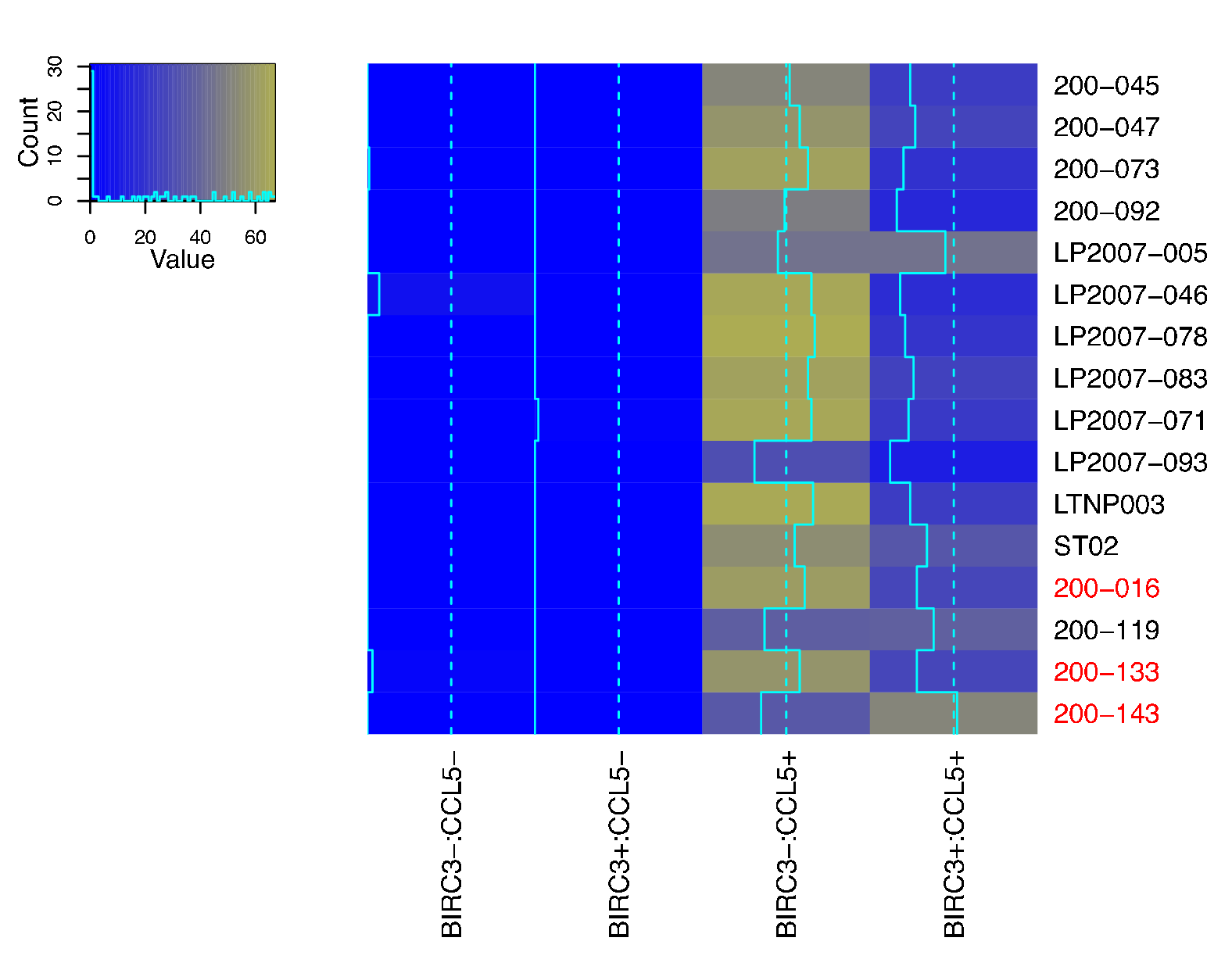}};
\begin{scope} [x={(foo.south east)},y={(foo.north west)}]
\node at (0,1) [font=\tiny\sffamily] {A} ;
\end{scope}
\end{tikzpicture}
};
\node [right=of A] (B){
\begin{tikzpicture}
\node[anchor=south west, inner sep=0] at (0,0) (bar) {\includegraphics[width=0.4\columnwidth]{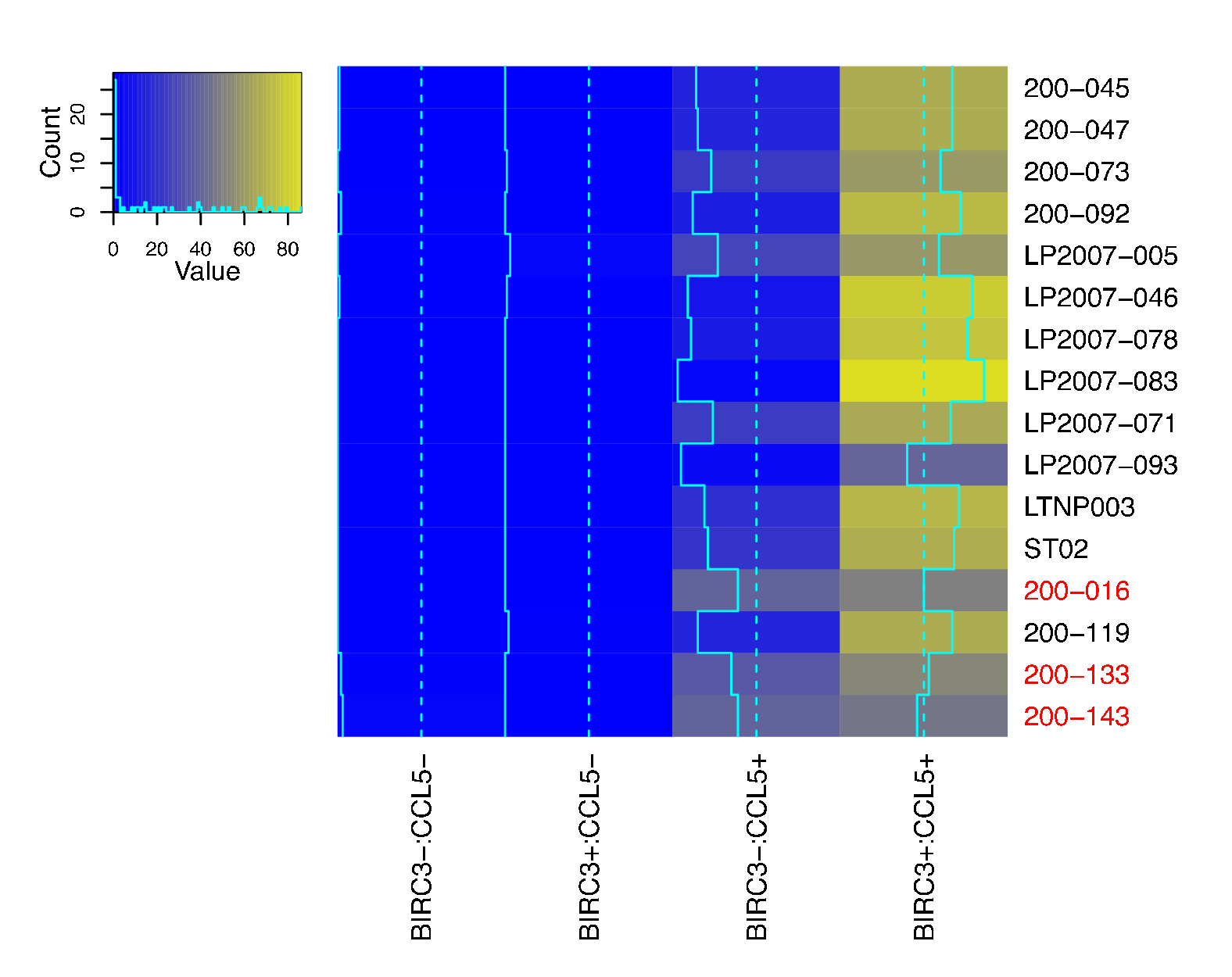}};
\begin{scope} [x={(bar.south east)},y={(bar.north west)}]
\node at (0,1) [font=\tiny\sffamily] {B} ;
\end{scope}
\end{tikzpicture}
};
%\node [anchor = south west, inner sep=0] at (0,-4) {\includegraphics[width=0.4\columnwidth]{Fluidigm_Multivariate_Marginalized_BIRC3_CCL5.pdf}};
%\node [anchor=south west, inner sep=0] at (6.5,-4) {\includegraphics[width=0.4\columnwidth]{Fluidigm_Multivariate_BIRC3_CCL5.pdf}};
%\node at (0,0) [font=\small\sffamily] {C} ;
%\node at (6.25,0) [font=\small\sffamily] {D} ;
\end{tikzpicture}
\caption{Counts of cells expressing different combinations of BIRC3 and CCL5 genes in the A) unstimulated and B) stimulated conditions. No difference is observed from the marginalized counts, while multivariate MIMOSA detects a difference between stimulated and unstimulated conditions in 13 of 16 samples. Sample names highlighted in red identify those where MIMOSA did not detect a difference. This figure appears in color in the electronic version of this article.}
\label{fig:polyfunctionality}
\end{figure}
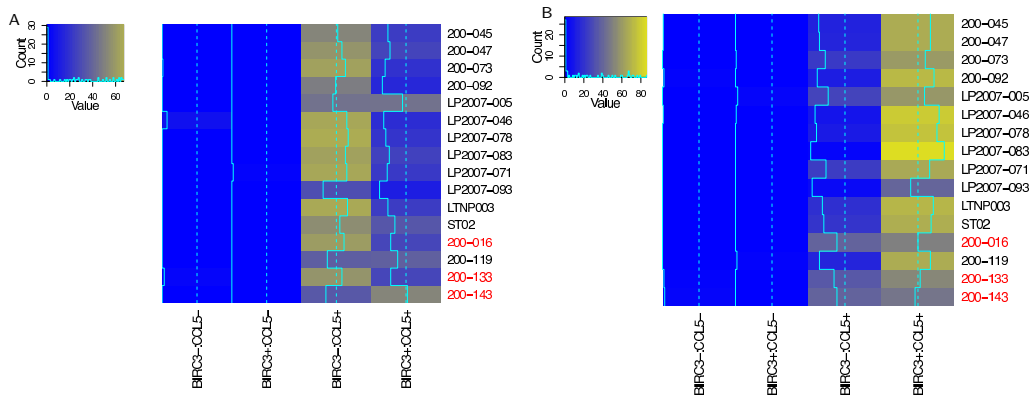

As a proof-of-concept, we applied our multivariate MIMOSA model for two specific genes in the Fluidigm data, namely BIRC3 and CCL5. For this example, $K=2$, and we have four possible combinations. 
In Figure~\ref{fig:polyfunctionality} we show heatmaps of the counts of cells expressing all combinations of the BIRC3 and CCL5 genes in unstimulated and stimulated samples (Figure~\ref{fig:polyfunctionality} A,B). Only CCL5 positive cells express BIRC3, and its expression increases upon stimulation. The typical approach to analyzing poly-functional populations from intracellular cytokine staining data (summing the counts over all possible polyfunctional cell populations  as in IL2$+$ and/or IFN-$\gamma+$) would not be appropriate in this case, since changes in the counts of these different cell populations occur in both directions. That is, the number of BIRC3-/CCL5+ cells decreases upon stimulation, while the number of BIRC3+/CCL5+ cells increases. When marginalizing over these cell populations, no difference is apparent in any of the samples. In contrast, the multivariate MIMOSA model tests all polyfunctional cell subpopulations simultaneously, and detects significant differences between stimulated and unstimulated conditions in 13 of the 16 samples (Figure~\ref{fig:polyfunctionality} D, black labels). Testing all combinations simultaneously is an advantage over performing multiple univariate tests on the subject combinations, which requires multiplicity adjustment and a potential loss of power. 

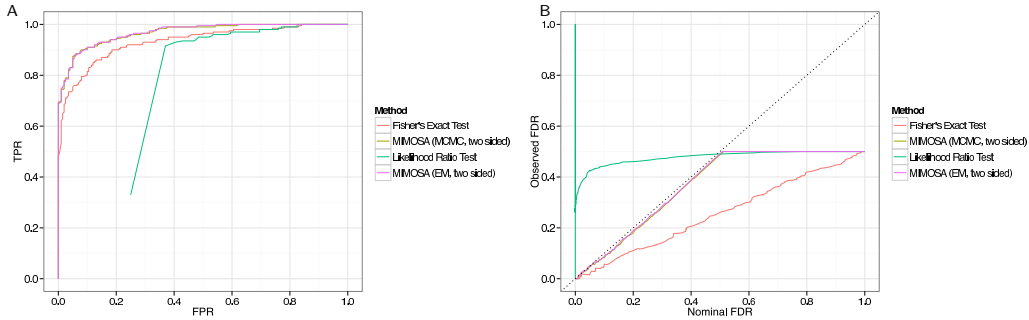
\begin{figure} %  figure placement: here, top, bottom, or page
   \centering
\begin{tikzpicture} [auto, node distance=0cm]
\node at (0,0) (A) {
\begin{tikzpicture}
    \node[anchor=south west, inner sep=0] at (0,0) (foo) {\includegraphics[width=0.4\columnwidth]{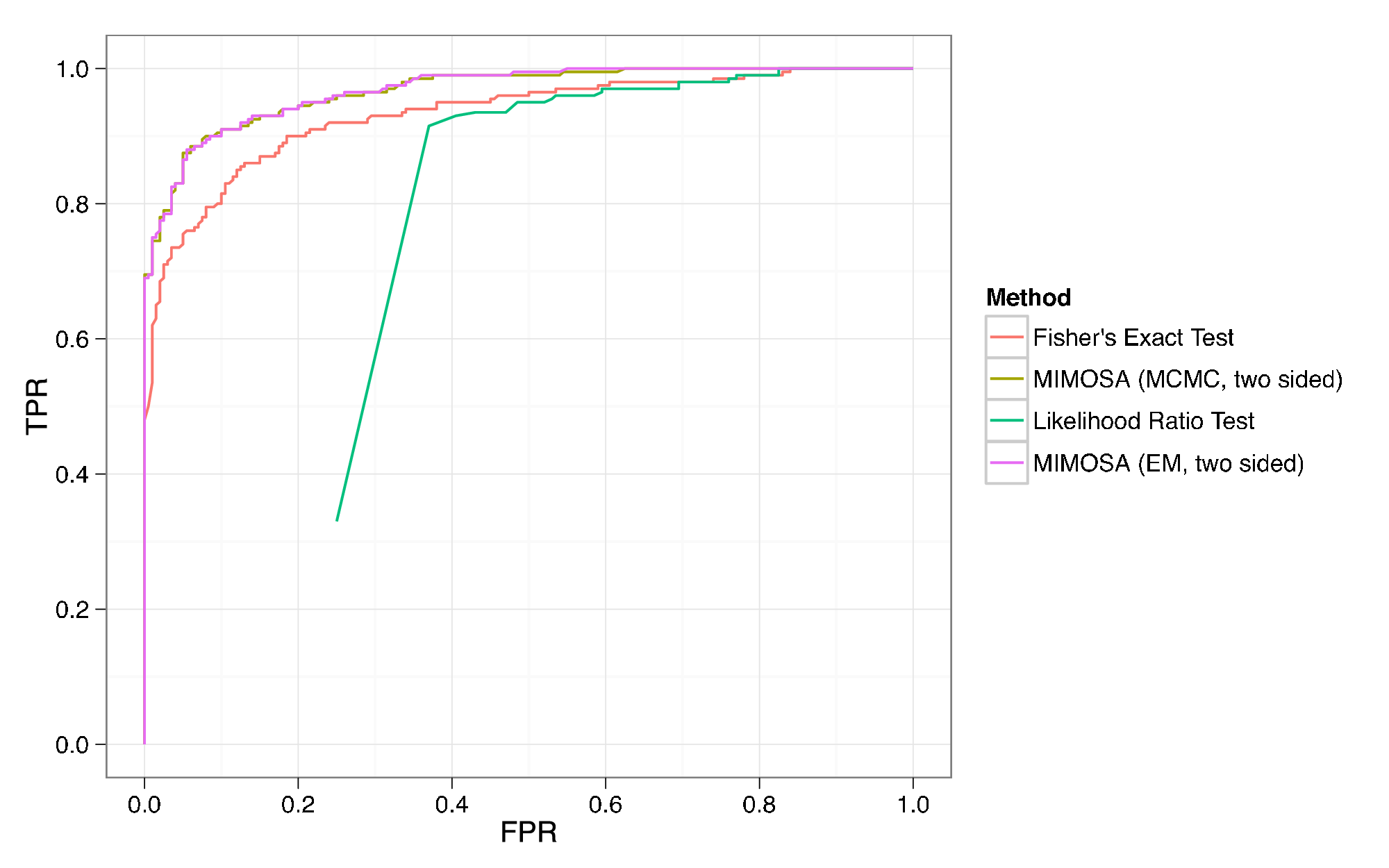}};
\begin{scope} [x={(foo.south east)},y={(foo.north west)}]
    \node at (0,1) [font=\tiny\sffamily] {A} ;
\end{scope}
\end{tikzpicture}
};
\node [right=of A] (B) {
\begin{tikzpicture}
   \node[anchor=south west, inner sep=0] at (0,0) (bar) {\includegraphics[width=0.4\columnwidth]{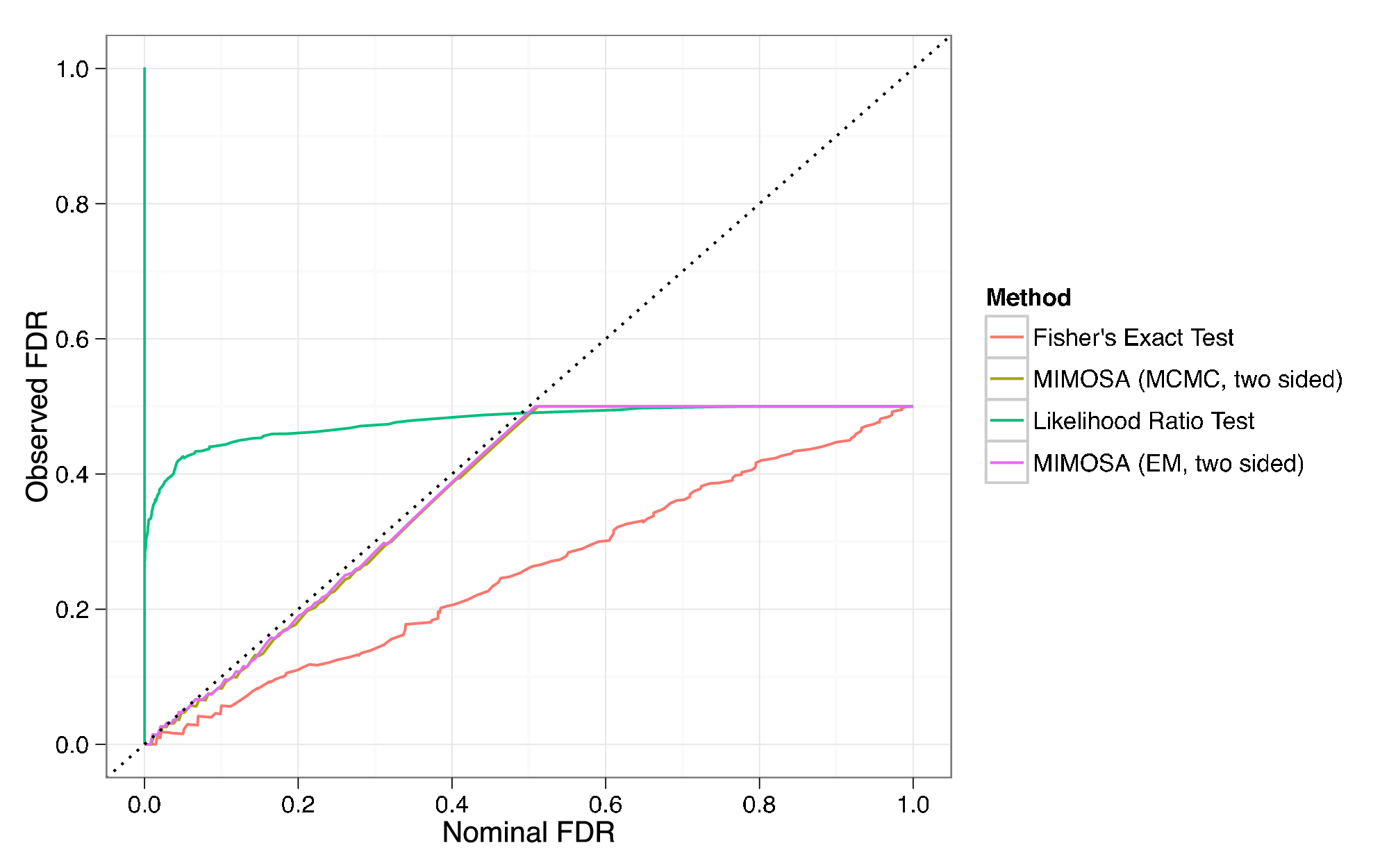}};
   \begin{scope} [x={(bar.south east)},y={(bar.north west)}]
       \node at (0,1) [font=\tiny\sffamily] {B} ;
\end{scope}
\end{tikzpicture}
};
\end{tikzpicture}
   \caption{Multivariate simulations from a two-sided model. Ten, eight-dimensional data sets were simulated from a two-sided model with an effect sizes of $2.5\times 10^{-3}$ and $-2.5\times 10^{-3}$ in two of the eight dimensions (N=1,500). Multivariate MIMOSA was compared against Fisher's exact test, and the likelihood ratio test. A) Average ROC curves for the competing methods over 10 simulations. B) Average observed and nominal false discovery rate for each method over 10 simulations. This figure appears in color in the electronic version of this article.}
   \label{fig:mvsimulations}
\end{figure}

Since the Fluidigm data set has a limited number of observations (100 cells and 16 samples), we could not look at more than two biomarkers at once. Therefore, we performed simulations in eight dimensions to assess the power of the multivariate MIMOSA  model compared to Fisher's exact test and the likelihood ratio test on the resulting 2x8 tables (Figure~\ref{fig:mvsimulations} A-C). These results show that multivariate MIMOSA has significantly increased power to detect true differences in multivariate data, even with small counts and small effect sizes, and the model better fits the data than the competing standard approaches tested (Figure~\ref{fig:mvsimulations} B).

\section{Discussion}
\label{s:discussion}
Experimentalists already have access to a myriad of single-cell assays such as flow cytometry, mass cytometry and multiplexed quantitative-PCR, to name a few. As single-cell assays become even more routine once sequencing at the single-cell level becomes practical~\citep{Ramskold:2012gj}, the development of effective statistical methods to detect differences in gene or protein expression at the single-cell level is becoming increasingly important. Current approaches for single-cell assays are for the most part simplistic such as the t-test, $\chi^2$ test, and Fisher's exact test, and resulting inference can be quite sub-optimal, especially when the cell counts are small. Most importantly, these methods do not share information across samples, resulting in less power to detect true differences than empirical-Bayes and hierarchical modeling approaches, which are widely applied in the microarray literature~\citep{Kendziorski:2003uw,Newton:2001go,Smyth:2005iy}. In addition, most of these methods are univariate in nature and inappropriate for high--dimensional, next--generation single-cell assays.

The MIMOSA model presented here uses a mixture model framework of beta-binomial or Dirichlet-multinomial distributions to model counts in experimental subjects across multiple conditions (\textit{i.e.}, vaccine responders and non-responders). Information is shared across responders and non-responders through exchangeable beta or Dirichlet priors, increasing the power to detect true differences between treatment and control conditions compared to Fisher's exact test, even when the underlying model assumptions are violated (Figures~\ref{fig:simulations} and Web Figure B). The univariate MIMOSA model based on the Beta-Binomial distribution allows us to constrain the alternative hypothesis to the case $p_s > p_u$, where the proportion of cells in the stimulated sample is strictly greater than the proportion of cells in the matched unstimulated sample. This has proven to be useful for the ICS data where stimulation induced changes are expected to be one-sided.

%Importantly, the analysis of real-world ICS data from vaccine trials demonstrated that the the constrained MIMOSA model performs as well or better than Fisher's exact test or other non-Bayesian alternatives for identifying vaccine responders across multiple antigen stimulations and multiple cytokines (Figure~\ref{fig:HVTN065}). Although the MIMOSA model was fit within each antigen stimulation $\times$ cytokine combination, the model is naive to vaccine time point. Despite this, MIMOSA demonstrated a higher sensitivity and specificity to discriminate between vaccine responders and non-responders on days 0 (pre-vaccination) and 182 (post-vaccination) than the current standard approach (Fisher's exact test), or non-Bayesian alternatives such as the likelihood ratio test or ranking by log-fold change (Figure~\ref{fig:HVTN065} A,C). Our approach treats all day 0 observations as true negatives and all day 182 observations as true positives, yet we know that not all vaccine recipients are likely to exhibit a vaccine response on day 182. None the less, this shortcoming, at worst, leads us to underestimate the true sensitivity of MIMOSA, and does not negatively impact the comparison. Additionally, MIMOSA was found to be a better fit to real-world ICS data from vaccine trials than other analysis approaches, as evidenced by more accurate estimates of true false discovery rate (Figure~\ref{fig:HVTN065} B,D).

Although we used two single-cell assay platforms as motivating examples, our MIMOSA model can be applied to any type of single-cell assay where cells are dichotomized into positive and negative sets, counted and compared across different conditions. 
In the case of the Fluidigm data, most analysis methods have been focused on identifying differences in the continuous part of the signal ignoring cells that are undetected (\textit{i.e.}, the gene is not expressed in the cell), or the information is used for pre-filtering~\citep{Flatz:2011jb}. 
The ability of MIMOSA to identify stimulation-specific expression patterns in single-cell gene expression data demonstrates not only the broader utility of the method, but importantly, also demonstrates that biologically relevant signal is present in the proportion of cells expressing each gene under different conditions (Figure~\ref{fig:fluidigm} A-C). 

Detecting differences in poly-functional cell populations (\textit{i.e.}, identifying changes in cell populations that co-express multiple proteins, cytokines, or genes) is important in immunology, since it allows the identification of more precisely defined, more homogeneous cell populations~\citep{Milush:2009bz}. 
In the context of HIV, poly-functional cell populations have been shown to be correlated with long-term disease non-progression, while in the context of vaccination studies (e.g. in Leishmania) poly--functional responses have been correlated with protection from disease~\citep{Betts:2006dw,Darrah:2007ih,Precopio:2007ht}. 
In the ICS data used here, the stimulation is expected to increase only the number of antigen specific cells detected.
Hence, if a specific cell subset expressing multiple biomarkers is being differentially expressed, differential expression based on the marginal cell counts should also be detected. 
As such, identifying poly-functional cytokine profiles from ICS data can be done in an iterative way. 
First, univariate tests on marginal populations are performed, and then specific cell subsets expressing the positive biomarkers detected are tested. 
However, this iterative (univariate) approach might not be satisfactory due to the large number of possible combinations that need to be tested, and a multivariate approach might be preferable. 
In that case,  as others have pointed out, in order to have the most power to detect a true difference, the statistical test should be selected taking into account only the cytokine combinations of interest~\citep{Nason:2006dx}. 

For two--sided changes, as with the Fluidigm data, changes in poly-functional cell populations are not always detectable when looking at the marginal populations (Figure~\ref{fig:polyfunctionality} A-C). 
In this case, the use of multivariate model, as our Dirichlet-multinomial model, will become important to detect differential biomarker expression. 
Here, we have shown that MIMOSA has higher sensitivity and specificity than the competing methods to identify true differences between conditions in multivariate count data (Figure~\ref{fig:polyfunctionality} A, and Figure~\ref{fig:mvsimulations} A,C), and the model generally provides a better fit to the single-cell assay count data obtained from studies with these types of experimental designs (Figure~\ref{fig:mvsimulations} B). 
Unfortunately, the limited number of samples in the Fluidigm data prevented us from looking at co-expression involving more than two genes. 
In the case of more than two biomarkers, the number of parameters to estimate for our Dirichlet--multinomial model is $2^{K+1}+1$, which is large even for moderate values of K. As an example, we would need both, a large number of subjects and a large number of events (cells) collected, to properly estimate the 33 parameters for $K=4$. %In the case of more than two markers, there is a combinatorial explosion in the number of parameters involved in our multinomial-Dirichlet mixture model. The number of parameters to be estimated is $2^{K+1}+1$, which can become very large even for moderate values of $K$. As an example, $K=4$ leads to 33 parameters, and this would require one to have a large number of subjects to properly estimate all parameters. 
A solution would be to explore alternative model parameterizations that could be used to reduce the number of required parameters. 
For example, one could assume that the hyper-parameters are constant across biomarker combinations, \textit{i.e.}, $\alpha_{\{s,u\}k}=\alpha_{\{s,u\}}$ for all $k$, and the number of parameters would be reduced to $3$ for any $K$. 
As attractive as this might sound, such a model would be unrealistic given that certain stimulations are known to induce expression of certain biomarkers more than others. 
More exploratory work will need to be done in this area once high dimensional single-cell level data with large number of samples become available.

All of the results presented here were obtained with a software implementation of the EM and MCMC MIMOSA models in R and C++, and is freely available from GitHub \\
(\url{http://www.github.org/gfinak/MIMOSA}). An R package will soon be released as part of the Bioconductor project~\citep{Gentleman:2004tt}.

\section*{Supplementary Materials}

Web Appendices A, B, C, and Web Figures A and B referenced in Sections \ref{s:data} and \ref{s:DEone},\ref{s:estimation},  and \ref{s:demarkercombos} are available in the attached Web-based supplementary material.\vspace*{-8pt}
% This will be replaced at submission time
%at the Biometrics website on Wiley Online Library.

\section*{Acknowledgments}
This work was supported by the Intramural Research Program of the  National Institute of Allergy and Infectious Diseases (NIAID) and the National Institutes of Health (NIH), and by grants R01 EB008400 and U01 AI068635-01 to RG, grants \#OPP38744, and \#OPP1032317 from the Bill \& Melinda Gates Foundation to VISC (Vaccine Immunology Statistical Center), grant \#OPP1032325 from the Bill \& Melinda Gates Foundation to the CAVD (Collaboration for AIDS Vaccine Discovery). Study HVTN065 was conducted by the HIV Vaccine Trials Network (HVTN), and supported by the National Institute of Allergy and Infectious Diseases (NIAID). The the HVTN Laboratory Program is supported by grant \#UM1AI068618. Funding was also provided by Public Health Service grant UM1 AI068618 from the NIH and the University of Washington Center for AIDS Research (CfAR), an NIH-funded program (P30 AI027757). We thank the James B. Pendleton Charitable Trust for their generous equipment donation.
\bibliographystyle{biom}
\bibliography{MIMOSA}
\label{lastpage}

\end{document}